\documentclass[10pt,aps,prl,reprint,superscriptaddress,preprintnumbers,nofootinbib,longbibliography,floatfix]{revtex4-2}

\usepackage[utf8]{inputenc}
\usepackage[T1]{fontenc}
\usepackage{anyfontsize}
\usepackage{microtype}

\usepackage{amsmath,amssymb,amsfonts,amsthm}
\usepackage{mathtools}
\usepackage{mathrsfs}
\usepackage{bm}
\usepackage{commath}
\usepackage{tensor}
\usepackage{leftindex}
\usepackage{braket}
\usepackage{accents}
\usepackage{cancel}
\usepackage{siunitx}
\usepackage{slashed}
\usepackage{graphicx}
\usepackage[dvipsnames]{xcolor}

\usepackage{array}
\usepackage{booktabs}
\usepackage{hhline}
\usepackage{tabularray}

\usepackage{enumitem}
\usepackage{csquotes}
\usepackage{chngcntr}
\usepackage{pifont}
\usepackage[normalem]{ulem}

\usepackage{comment}
\usepackage{verbatim}
\usepackage{lipsum} 

\usepackage{xargs} 

\usepackage{xcolor}
\definecolor{darkcandyapplered}{rgb}{0.64, 0.0, 0.0}
\definecolor{azure(colorwheel)}{rgb}{0.0, 0.5, 1.0}
\definecolor{darkcerulean}{rgb}{0.03, 0.27, 0.49}
\definecolor{darkcyan}{rgb}{0.0, 0.55, 0.55}
\definecolor{darkolivegreen}{rgb}{0.33, 0.42, 0.18}
\definecolor{armygreen}{rgb}{0.29, 0.33, 0.13}
\definecolor{myrtle}{rgb}{0.13, 0.26, 0.12}
\usepackage[
	colorlinks=true,
	linkcolor=darkcerulean,
	citecolor=darkcandyapplered,
	urlcolor=myrtle,
	filecolor=blue,
	pdfstartview=FitV,
	bookmarksopen=true
]{hyperref}

\usepackage{bibunits}

\defaultbibliographystyle{custom_apsrev4-2}
\usepackage{orcidlink}
\usepackage{xurl}

\DeclareMathAlphabet{\mathpzc}{OT1}{pzc}{m}{it}

\makeatletter
\if@twocolumn

\else

\fi
\makeatother

\renewcommand{\Re}{\operatorname{Re}}
\renewcommand{\Im}{\operatorname{Im}}

\newcommand{\be}{\begin{equation}}
\newcommand{\ee}{\end{equation}}

\newcommand{\ba}{\begin{eqnarray}}
\newcommand{\ea}{\end{eqnarray}}

\newcounter{suppsec}
\newcommand{\suppsection}[1]{%
  \refstepcounter{suppsec}%
  \section*{\Roman{suppsec}.\ #1}%
}

\newcounter{suppsubsec}[suppsec]

\newcommand{\IITGN}{Department of Physics, Indian Institute of Technology Gandhinagar, Palaj, Gandhinagar 382055, Gujarat, India}

\newcommand{\rBH}{r^{\rm BH}}          
\newcommand{\tBH}{t^{\rm BH}}          
\newcommand{\RBH}{\mathcal{R}^{\rm BH}}  
\newcommand{\Gh}{\hat{\mathcal G}}
\newcommand{\sh}{\hat\sigma}
\newcommand{\R}{\mathcal{R}_s} 

\newcommand{\Mf}{M_{\rm f}}
\providecommand{\lnB}{\ln\mathcal{B}}

\begin{document}
	
	\title{Listening to the Horizon: Probing Near-Horizon Reflectivity with GW250114}

    \author{R. Prasad~\orcidlink{0000-0002-6602-3913}}
	\email{prasad@iitgn.ac.in}
	\affiliation{\IITGN}
    
	\author{Subhodeep Sarkar~\orcidlink{0000-0001-9015-8837}}
	\email{ssarkar@iitgn.ac.in}
	\affiliation{\IITGN}

     \author{Sudipta Sarkar~\orcidlink{0000-0002-4077-1358}}
	\email{sudiptas@iitgn.ac.in}
	\affiliation{\IITGN}
	
	\date{07 October 2026}
	
	\begin{abstract}
Testing the horizon properties of astrophysical black holes is a central goal of gravitational-wave astronomy. The recently identified Direct Wave (DW), a prompt non-quasinormal component of GW250114, provides a new probe of near-horizon physics. While it remains unclear whether the DW itself carries horizon information, we show that it enables a novel null test of horizon reflectivity. If the remnant is an exotic compact object (ECO) with a partially reflecting surface, inward plunge radiation is reflected and leaks back through the gravitational potential barrier, producing a delayed DW echo whose amplitude is set by the surface reflectivity ($\R$). We construct the corresponding echo template, normalized by the observed DW, and search for it in quasinormal-mode-filtered data. We find no evidence for an echo ($\ln\mathcal{B}=-2.9$) and obtain the 90\% credible bound $|\R|<0.06$, corresponding to a reflected energy fraction below $0.4\%$, for surfaces from $0.1\, \Mf$ to a Planck length from the would-be horizon. Our method thus turns the prompt ringdown signal into a direct test of horizon reflectivity.
\end{abstract}
    
	\maketitle
    
\begin{bibunit}

\emph{\textbf{Introduction.---}}The event horizon is the defining feature of a black hole (BH), yet observational access to the near-horizon region remains indirect. Electromagnetic observations probe this region through the shadow and photon ring, the inner accretion flow, or jet launching. Gravitational waves (GWs) provide a potentially cleaner route, with BH spectroscopy~\cite{Berti:2009kk,Berti:2025hly,Cardoso:2025npr} emerging as the standard framework for testing the merger remnant \cite{LIGOScientific:2025wao, LIGOScientific:2026oim}.

BH spectroscopy exploits quasinormal modes (QNMs), whose complex frequencies in general relativity are determined entirely by the remnant mass and spin. Measuring multiple QNMs therefore enables stringent tests of the Kerr hypothesis \cite{LIGOScientific:2026oim, Krishnendu:2021fga}, turning observational BH physics into a precision science.

However, the initial ringdown does not directly probe the horizon~\cite{Cardoso:2016rao}. Although QNM frequencies carry signatures of the compact object and perhaps its environment~\cite{Cheung:2021bol,Jaramillo:2020tuu,Jaramillo:2021tmt}, the prompt ringdown is controlled mainly by the effective potential near the light ring. A ringdown consistent with a BH spacetime therefore constrains the geometry around the photon surface while remaining comparatively insensitive to modifications near the horizon at $r=r_+$~\cite{NotationConventions}. Such structure can instead appear at later times as delayed repetitions of the initial perturbation~\cite{Cardoso:2017njb,Cardoso:2017cqb,Cardoso:2019rvt}. Despite initial optimism~\cite{Abedi:2016hgu,Conklin:2017lwb,Abedi:2018npz}, searches for these late-time \emph{echoes} have found no statistically significant evidence~\cite{Westerweck:2017hus,Lo:2018sep,Tsang:2019zra,Miani:2023mgl,Uchikata:2023zcu,Wu:2025enn}. These searches either leave the echo amplitude free, so that a null result constrains a template amplitude rather than a surface property, or tie it to the observed ringdown by reflecting it off the potential barrier~\cite{Wang:2019rcf,Maggio:2019zyn}. What remains missing is an independent prediction for the radiation reaching the horizon.

\begin{figure}[htbp!]
\centering
\includegraphics[width=\columnwidth]{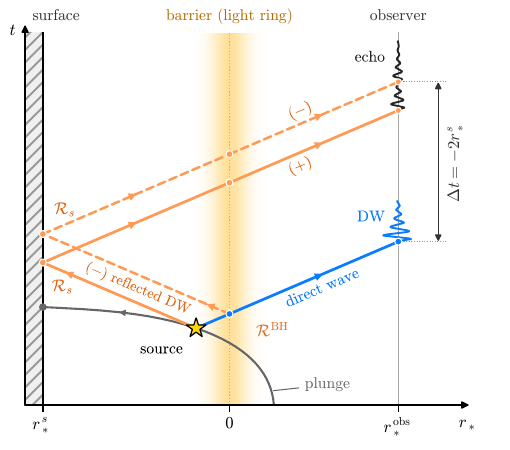}
\caption{Schematic spacetime picture of the first echo. Radiation from the plunging source (star) reaches the observer through the barrier as the DW (blue). Radiation emitted toward the horizon, $(+)$, and the part of the outgoing radiation reflected by the barrier, $(-)$ ($\RBH$), are reflected at the surface ($\R$) and cross the barrier outward as the echo. The barrier modulates their frequency content. The $(-)$ path is delayed by exactly $\Delta t=-2r_*^s$ with respect to the DW.}
\label{fig:schematic}
\end{figure}

Recently, a qualitatively distinct, non-QNM component of the post-peak signal has been identified. Oshita \emph{et al.}~\cite{Oshita:2025qmn} showed that the prompt emission generated during the plunge can contribute substantially to the merger waveform, identifying this component as the \textit{Direct Wave} (DW). Lu \emph{et al.}~\cite{Lu:2025vol} subsequently reported detecting the DW in GW250114, the loudest binary-BH merger observed to date~\cite{LIGOScientific:2025rid}, presenting it as an observational channel for frame dragging and near-horizon redshift.

The idea of Direct Waves may be traced back to an early work~\cite{Mino:2008at} showing that a plunging source generates modes distinct from QNMs, with frequencies depending entirely on the Kerr parameters. A subsequent analysis~\cite{Zimmerman:2011dx} corrected some of these initial results, establishing that these \emph{horizon modes} are generic and suggesting that some of these modes might leave an imprint on the ringdown. However, Ref.~\cite{Oshita:2025qmn} argues that the entire family of horizon modes, {$\omega_{H,n} = m\Omega_H-i n \kappa$}, is screened (where $\Omega_H$ and $\kappa$ are the angular velocity and surface gravity of the remnant BH). Nevertheless, it was hypothesized that a connection could still exist: the Direct Wave carries a time-dependent frequency $\omega_G(t)$ set by the source, which approaches (but never coincides with) $\omega_{H,n}$, despite being generated primarily near the light ring.

By removing the dominant QNMs using frequency-domain QNM filters~\cite{Ma:2022wpv}, a coherent residual consistent with the DW was recently identified in observational data, exhibiting signal-to-noise ratios (SNRs) of $15.8^{+0.1}_{-0.5}$ and $17.1^{+0.1}_{-0.4}$ in the Hanford and Livingston data~\cite{Lu:2025vol}. This new DW component has also been corroborated by independent time-domain subtraction methods~\cite{Chung:2026eph} and its identification in numerical-relativity (NR) simulations~\cite{Dyer:2026yex}.

The extent to which the DW itself probes horizon-scale geometry remains unsettled~\cite{Kankani:2026byb,Kankani:2026kst,Dyer:2026yex,Kuntz:2026xep,Ma:2026hcb}. Both fixed- and free-frequency analyses indicate a drifting DW frequency that does not simply track the horizon modes~\cite{Kankani:2026byb,Kankani:2026kst,Dyer:2026yex}. Moreover, calculations locate the dominant observable DW emission closer to the light-ring region than to the horizon~\cite{Oshita:2025qmn,Cheung:2026gfd}. \textit{Our analysis, however, does not require the DW to be a horizon mode}. We use it only as an empirically calibrated template for a distinct post-peak component demonstrably present in the data. Our constraints therefore do not depend on interpreting its instantaneous frequency in terms of $(\Omega_H,\kappa)$.

What makes the DW particularly useful is its temporal and spectral separation from the conventional ringdown. Localized near the merger peak and spectrally distinct from the QNMs, it remains the strongest coherent structure after subtraction of the dominant modes. Within the point-particle plunge model, the DW therefore calibrates the radiation sent toward the horizon. A partially reflecting surface returns a fraction of this radiation after a location-dependent delay; the resulting first echo is not a delayed copy of the DW, since the barrier transmits mainly its high-frequency component. The theoretical basis is developed in the End Matter.

In this Letter, we apply this test to GW250114~\cite{LIGOScientific:2025rid,Lu:2025vol}. We compute the first plunge echo from the Teukolsky equation with an inner boundary of reflectivity $\R$ and search over surface locations from $0.1\Mf$ to a Planck length $\ell_p$ from the horizon. Using a single event and a single round trip, the test complements spectroscopy: QNMs probe the exterior Kerr spacetime, while the echo probes the response of its innermost boundary.

\emph{\textbf{Direct Wave.---}}{Following Refs.~\cite{Oshita:2025qmn, Lu:2025vol}, we model} the DW contribution to the $(\ell,m)=(2,2)$ strain mode as
\begin{equation}
h_{\rm DW}(t)=\bigl[\omega_G(t)-\omega_H\bigr]\,e^{-i\int^{t}\omega_G(t')\,dt'},
\label{eq:dw}
\end{equation}
where $\omega_G(t)=2\hat{\Omega}(t)-i\hat{\kappa}(t)$ {and the horizon mode is  taken to be $\omega_{H} \equiv \omega_{H,n=1} = 2 \Omega_H - i \kappa$}, with
\begin{equation}
\hat{\Omega}=(\beta\Omega_H+\dot{\phi})/(1+\beta),\quad
\hat{\kappa}=2\beta\kappa/(1+\beta),\quad
\label{eq:omegaG}
\end{equation}
and $\beta\equiv-\dot r_*$. These quantities are evaluated along a plunging trajectory in the Kerr spacetime of the remnant, so the shape of $h_{\rm DW}$ is fixed, with no free shape parameters. In the analysis it enters as $A\,h_{\rm DW}(t-\delta t)$, where the complex amplitude $A$ sets the overall normalization and phase, and $\delta t$ aligns the template with the data. For GW250114, our posterior on $(|A|,\arg A,\delta t)$ is consistent with Ref.~\cite{Lu:2025vol} (Supplemental Material, SM~\cite{SMDW}).

 \smallskip
\emph{\textbf{The first echo of the plunge.---}}We place a (partially) reflecting surface at $r_s= r_{+} \,+  \,\epsilon$, a small coordinate distance $\epsilon \ll M$ outside the would-be horizon, corresponding to the tortoise coordinate $r_*^s<0$, with flux-normalized reflectivity $\R=|\R|e^{i\phi_{R}}$, taken to be constant across the analysis band~(see End Matter~\cite{Chen:2020htz,Xin:2021zir,Ma:2022wpv}). The region between the surface and the potential barrier forms a cavity with round-trip time~\cite{Abedi:2016hgu}
\begin{equation}
\Delta t=-2r_*^s\simeq-\bigl[4Mr_+/(r_+-r_-)\bigr]\ln(\epsilon/2M).
\label{eq:dt}
\end{equation}
We quote the position of the surface by its proper distance from the horizon, $d_s\simeq2r_+[\epsilon/(r_+-r_-)]^{1/2}$ in the equatorial plane. Note that the radial Teukolsky equation admits two homogeneous solutions~\cite{Mino:2008at, SMDW}, $R^{\rm in}$, ingoing at the horizon, and $R^{\rm up}$, outgoing at infinity.

For a BH, the source generates the amplitudes $Z^\infty$ and $Z^H$ of the radiation escaping to infinity and entering the horizon, respectively; each is a source integral over the homogeneous solution regular at the opposite boundary. With the horizon replaced by a reflecting surface (End Matter), the amplitude at infinity becomes a sum over successive round trips, whose first term is the first echo,
\begin{equation}
Z^{(1)}=\Gh\,\tBH\,Z^H,\qquad \Gh=\sh\,\R\,e^{ik_H\Delta t}.
\label{eq:Z1}
\end{equation}
Here $\tBH=C^{\rm trans}/C^{\rm up}$ is the transmission amplitude of the barrier for a wave propagating outward from the horizon, and $\sh=(-1)^{m+1}\sigma/4$, with $\sigma$ the Teukolsky--Starobinsky factor~\cite{Chen:2020htz,Xin:2021zir,Ma:2022wpv,SMDW}. Equation~\eqref{eq:Z1} has a direct physical interpretation: the first echo is the radiation that would have been absorbed by a BH, returned by the surface, and transmitted back through the potential barrier. The barrier's flux-normalized reflectivity~\cite{Rosato:2026apq} is $\RBH=\sh\,\rBH$, with $\rBH=C^{\rm ref}/C^{\rm up}$; its transmissivity $\Gamma=1-|\RBH|^2$ sets the fraction that escapes: $|\Gamma|\ll1$ below the barrier peak, $\Gamma\approx0.5$ near the peak ($f\simeq250$\,Hz), and $\Gamma\to1$ above it.

The two horizon components of $R^{\rm up}$ split $Z^H=Z^H_{(-)}+Z^H_{(+)}$ (End Matter), and correspondingly decompose the first echo into the two propagation paths of Fig.~\ref{fig:schematic}:

\begin{equation}
Z^{(1)}\simeq\R\,e^{ik_H\Delta t}\bigl[\RBH\,Z^\infty+\sh\,\tBH\,Z^H_{(+)}\bigr].
\label{eq:twoparts}
\end{equation}

The two contributions differ in the sign of the near-horizon phase, $e^{\mp ik_Hr_*}$. The $(-)$ contribution is the outgoing wave, which contains the DW, reflected inward by the potential barrier and returned after the delay $\Delta t$; its frequency sweeps toward $m\Omega_H$ as the particle approaches the horizon. By contrast, for the $(+)$ contribution the saddle frequency diverges as the coordinate velocity $\beta\to1$, while the advanced time $v=t+r_*$ freezes, so the late-time emission is compressed into a short, broadband burst. {Because the $(+)$ contribution has no regular saddle point, it has no simple closed-form DW analogue  (End Matter), and we therefore compute the complete echo numerically.}

\smallskip 

\begin{figure}[htbp!]
\centering
\includegraphics[width=\columnwidth]{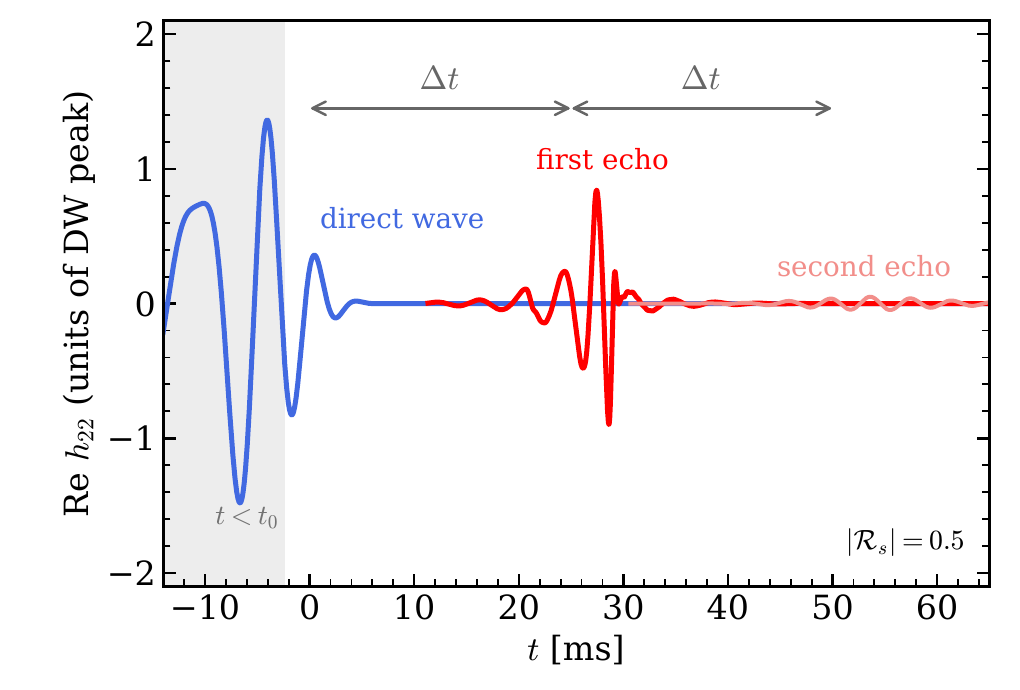}
\caption{QNM-filtered DW and first echo of the plunge, computed numerically from the Teukolsky equation for the GW250114 remnant, in units of the DW peak at $t\ge t_0$: DW (blue), first echo for $|\R|=0.5$ and $\Delta t=25$\,ms (red), and second echo (light red).  $t=0$ is the peak of the unfiltered waveform; grey region: $t<t_0=-7\Mf$.}
\label{fig:template}
\end{figure}

\emph{\textbf{Echo template.---}}We solve the radial Teukolsky equation for a point particle plunging equatorially into the GW250114 remnant ($\chi_f=0.6725$), on a trajectory calibrated to numerical relativity as in Ref.~\cite{Xin:2021zir}: up to the innermost stable circular orbit (ISCO) its orbital frequency follows that of an equal-mass SEOBNRv4 waveform~\cite{Bohe:2016gbl}, and after the ISCO it follows a plunging geodesic. The source is switched on $300\Mf$ before the light-ring crossing, so that the late inspiral and the complete plunge contribute. Near the horizon, where the numerical solution loses accuracy, we continue it with a Frobenius expansion and remove the non-radiative self-field feature of Ref.~\cite{Xin:2021zir}. Our waveform and echoes agree with theirs (SM Sec.~II). The template is
\begin{equation}
\begin{aligned}
\tilde h(\omega)&=A \,\big[\tilde h_{\rm DW}(\omega)+\R\,e^{ik_H\Delta t}\,\tilde h_{E1}(\omega)\big],\\
\tilde h_{E1}&\propto F(\omega)\,\sh\,\tBH\,Z^H/\omega^2,
\end{aligned}
\label{eq:template}
\end{equation}
where $F$ is the rational filter of Ref.~\cite{Ma:2022wpv}, which removes the $(2,2,n\le2)$ QNMs; it is applied alike to the data and the numerical templates, and is built from the Kerr QNMs of the remnant even for an ECO~\cite{Note_QNM_Filter}. In the search, $\tilde h_{\rm DW}$ is the analytic model of Eq.~\eqref{eq:dw}, which contains no QNMs and is not filtered; the QNM-filtered DW of the numerical source matches it (match $0.9993$ \cite{Owen:1995tm}, End Matter) and so sets the normalization and time origin of the numerical echo $\tilde h_{E1}$. Since the DW and the echo are emitted by the same plunge, the DW measures the complex amplitude $A$, which the echo shares: $\R$ is the only remaining parameter.
 
In Fig.~\ref{fig:template}, we show the echo in the time domain. Its spectrum carries most of its power above $\approx250$\,Hz (see Fig.~\ref{fig:S_spec}), where the barrier becomes transparent, and is not merely a delayed replica of the DW. Higher-order echoes carry further factors of $\R\RBH$ and are negligible for $|\R|\ll1$; therefore, neglecting them is conservative.
\begin{figure}[tb]
\centering
\includegraphics[width=\columnwidth]{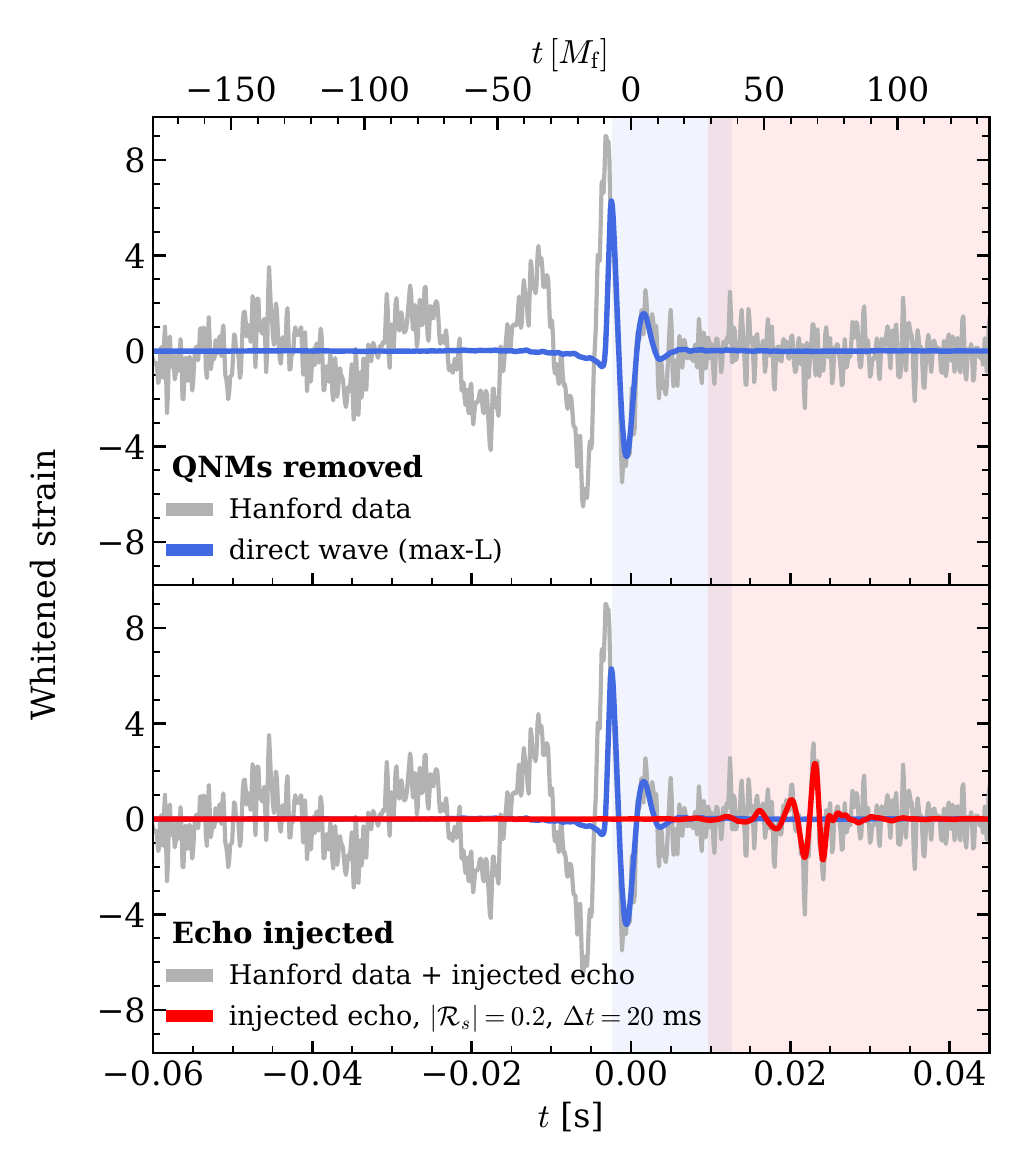}
\caption{Whitened, QNM-filtered Hanford data (grey). Top: the maximum-likelihood DW (blue, analytical model), which closely follows the data. Bottom: the data with an injected echo of $|\R|=0.2$ and $\Delta t=20$\,ms (red; network SNR $\approx11$). Data and templates are whitened with the broadband noise PSD and band-passed to 50--2000\,Hz. Shaded: DW region (blue) and echo-search window (red). Like the filtered data, the DW and the echo contain no QNMs.}
\label{fig:data}
\end{figure} 

\smallskip

\emph{\textbf{Search in GW250114.---}}We analyze the H1 and L1 data after removing the $(2,2,0)$, $(2,2,1)$ and $(2,2,2)$ QNMs at the maximum-posterior remnant parameters, following Ref.~\cite{Lu:2025vol}, with a time-domain Gaussian likelihood on a $0.32$\,s segment that starts at $t_0=-7\Mf$ from the strain peak (detector frame remnant mass $\Mf=67.9\,M_\odot$). We fit the DW and the echo together with Eq.~\eqref{eq:template}, whose parameters are $(A,\delta t,|\R|,\phi_R,\Delta t)$. At fixed $(\R,\Delta t,\delta t)$ the model is linear in $A$, which we marginalize analytically; $\delta t$ is marginalized on a grid, and the posterior of $(\R,\Delta t)$ is computed on a grid. The DW is thus neither subtracted nor held fixed, and its uncertainty propagates into the echo; alternative treatments give the same bound (End Matter). We adopt uniform priors $|\R|\in[0,1]$, $\phi_R\in[0,2\pi)$ and $\Delta t\in[12,300]$\,ms, whose lower end keeps the echo separated from the DW. 
Through Eq.~\eqref{eq:dt}, this delay range corresponds to proper distances $10^{-41}\lesssim d_s /\Mf\lesssim10^{-1}$, uniformly in $\ln d_s$, and includes a surface one Planck length from the horizon ($\Delta t\approx 292$\,ms). Figure \ref{fig:data} shows the QNM-removed Hanford data with the best-fit DW, which closely follows the data; an injected echo with $|\R|=0.2$, corresponding to a network SNR of $\approx11$, would be clearly visible.

\begin{figure}[ht]
\centering
\includegraphics[width=\columnwidth]{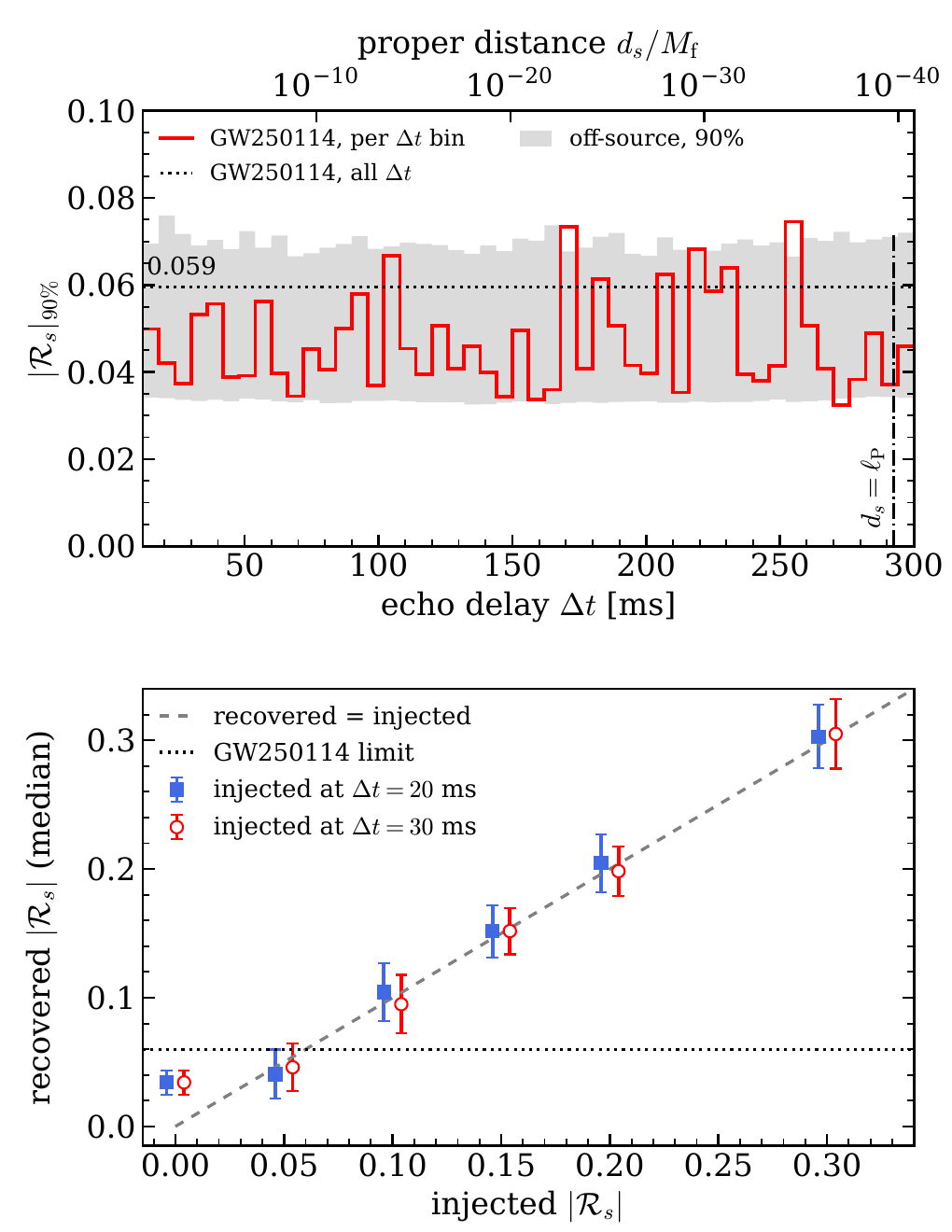}
\caption{Top: 90\% upper limit on $|\R|$ for GW250114 (red) in $6$\,ms bins of $\Delta t$, compared with the 5--95\% range expected from noise alone (grey); dotted: the limit over all $\Delta t$. The top axis gives the proper distance $d_s$ of the surface from the horizon, in units of $\Mf$. Bottom: recovered against injected $|\R|$ for echoes at $\Delta t=20$ and $30$\,ms, injected into $50$ off-source segments. Mean of the posterior medians (markers) and standard deviation over segments (bars) are shown. The injected values are recovered well.}
\label{fig:constraint}
\end{figure}
 
\emph{\textbf{Results.---}}We find no evidence for an echo. The Bayes factor of the echo model over the DW alone is $\ln\mathcal{B}=-2.9$. Marginalizing over the full delay range, we obtain the 90\% credible upper limit
\begin{equation}
|\R|<0.059\qquad(10^{-41}\lesssim d_s/\Mf\lesssim10^{-1}),
\label{eq:bound}
\end{equation}
that is, a reflected energy fraction $|\R|^2<3.5\times10^{-3}$, or approximately $0.4\%$ [Fig.~\ref{fig:constraint} (top), dotted line].  We also determine the limit as a function of delay by
repeating the analysis in $6$\,ms bins of $\Delta t$ [Fig.~\ref{fig:constraint} (top), red]. 

To calibrate against noise, we analyze off-source data: $0.32$\,s stretches of H1 and L1 data recorded $16$--$49$\,s after the merger, each with the maximum-likelihood DW added. Pairing the H1 data of one stretch with the L1 data of another yields $3000$ noise realizations, whose 5--95\% range (grey) is the expectation for noise alone. The GW250114 limit lies within this range; the few bins that fall outside it are consistent with chance, as discussed in SM Sec.~III.

The bound therefore holds across the entire range of $d_s$ searched, down to the Planck scale. We also carry out injection-recovery studies. For echoes injected into the off-source data, the recovered $|\R|$ tracks the injected value [Fig.~\ref{fig:constraint} (bottom)]: an echo with $|\R|=0.1$ (SNR $\approx6$) or louder would not have been missed, and the 90\% limit lies above the injected value in $92\%$ of random injections (SM Sec.~III). A reflecting surface with $|\R|\gtrsim0.06$ is therefore disfavored, and a perfect reflector, $|\R|=1$, is excluded.

\begin{figure}[htbp!]
\centering
\includegraphics[width=\columnwidth]
{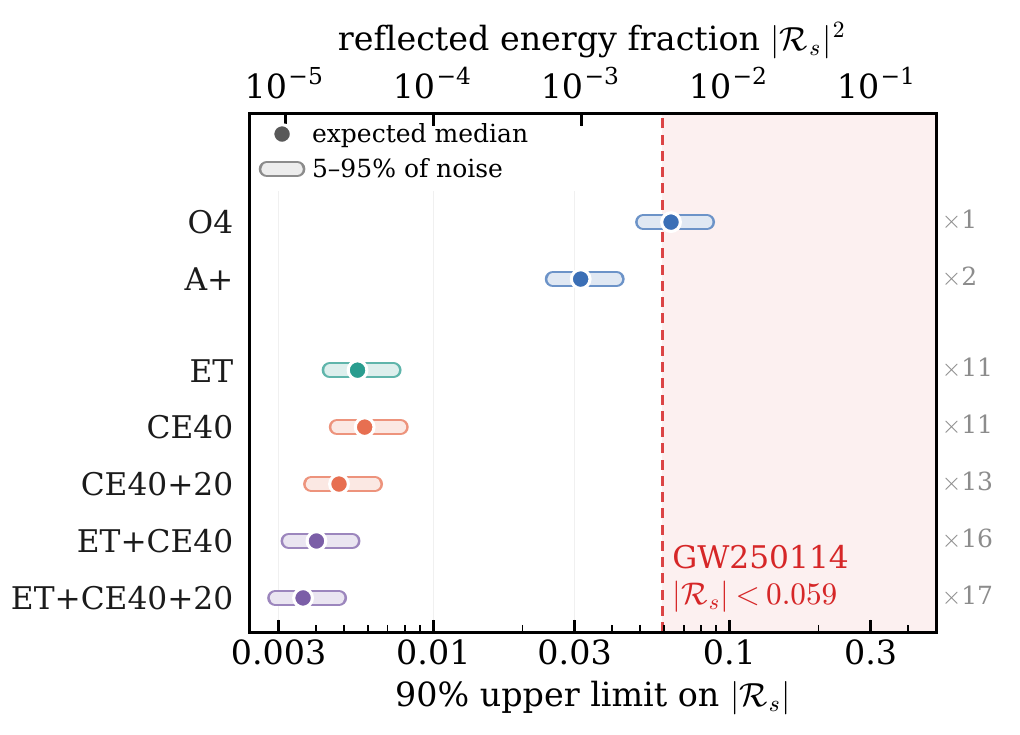}
\caption{Expected 90\% upper limits on $|\R|$ ($\Delta t$ marginalized) for a GW250114-like event in current and future detector networks. They are the limits found in O4 noise without an echo (off-source data; median $0.063$), divided by each network's gain in echo SNR over O4 (right axis). Circles: median; bars: 5--95\% range. Dashed line and shading: the GW250114 limit ($0.059$) and the values it excludes. Top axis: reflected energy fraction $|\R|^2$.}
\label{fig:next_generation_constraint}
\end{figure}

\emph{\textbf{Forecast.---}}Next-generation detectors will record GW250114-like signals routinely. Keeping the source fixed (distance, sky location, orientation), we compute the echo SNR per unit $|\R|$, $\rho_1$, with the design noise curves distributed with \texttt{gwbench}~\cite{Borhanian:2020ypi}, and divide the O4 off-source limits by the gain $\rho_1/\rho_1^{\rm O4}$ (Fig.~\ref{fig:next_generation_constraint}). This is a scaling estimate, not a search in simulated noise: in Gaussian noise the limit scales as $1/\rho_1$. LIGO A+~\cite{Barsotti:2018apl} would halve it; the Einstein Telescope (ET)~\cite{Punturo:2010zz,Maggiore:2019uih} or a single 40-km Cosmic Explorer (CE)~\cite{Reitze:2019iox,Evans:2021gyd} would tighten it tenfold, and ET with a 40-km and a 20-km CE by a factor of $\approx$ 17. A GW250114-like event in such a network would limit the energy returned by the surface to about one part in~$10^{5}$. 

\medskip

\emph{\textbf{Conclusions.---}}The defining property of an event horizon is not the spectrum of radiation outside a black hole, but what happens to radiation reaching its innermost region: a classical horizon absorbs it without return~\cite{Sarkar:2019xfd}. Testing this directly is difficult. BH spectroscopy mainly probes the exterior potential near the light ring~\cite{Cardoso:2016rao}, while conventional echo searches generally leave the returning amplitude free. The Direct Wave in GW250114 offers a different route: within our model, it calibrates the plunge radiation and hence fixes the normalization of the inward component, turning the absence of a return into a bound on near-horizon reflectivity.

We find no evidence for such a signal. For surface locations from $0.1\, \Mf$ down to a Planck proper distance from the would-be horizon, we obtain $|\mathcal{R}_s|<0.059$ at $90\%$ credibility, implying that less than $0.4\%$ of the incident energy is reflected. This directly constrains the inner-boundary response rather than a generic echo amplitude, using only the first return of the observed plunge radiation.

The main theoretical uncertainty is the point-particle plunge model, calibrated here to numerical relativity; a comparable-mass calculation will test the mapping between the observed DW and inward radiation. The framework can also accommodate frequency-dependent reflectivity and specific quantum-BH or ECO models \cite{Wang:2019rcf, Chakravarti:2021clm, Cardoso:2017njb}, including Boltzmann- or Lorentzian-type responses. With next-generation detectors, a GW250114-like event could constrain the reflected energy fraction to $\sim10^{-5}$, entering the range predicted by some quantum-modified horizon models. Direct Waves may therefore provide a direct measure of how nearly perfectly a merger remnant absorbs gravitational radiation.

A further assumption is the Kerr exterior. Although no regular interior is known to match an exact Kerr exterior, non-Kerr multipoles are expected to be suppressed as the ECO surface approaches the would-be horizon~\cite{Raposo:2018xkf,Barcelo:2019aif}. Nevertheless, extending the analysis to parametrized non-Kerr geometries, such as Hartle--Thorne or Manko--Novikov, would test the robustness of our constraints.

\smallskip
\begin{acknowledgments}
\emph{\textbf{Acknowledgments.---}}Sudipta Sarkar’s research is supported by the Anusandhan National Research Foundation, India, under the ARG Grant (No. ANRF/ARG/2025/005821/PS). This research has made use of data obtained from the Gravitational Wave Open Science Center, a service of the LIGO Scientific Collaboration, the Virgo Collaboration, and KAGRA.
\emph{Data availability.---} Notebooks and scripts for reproducing the results~\cite{GitHubRepoRP} will be made available in the accompanying GitHub repository upon publication (or on reasonable request). \emph{Tools and Software.---}We have used \texttt{pybhpt}~\cite{Nasipak_pybhpt_2026,Nasipak:2022xjh,Nasipak:2025tby}, \texttt{Black Hole Perturbation Toolkit}~\cite{BHPToolkit,Neef:2026qoq}, \texttt{qnmfilter}~\cite{Ma:2022wpv}, \texttt{bilby}~\cite{Ashton:2018jfp}, \texttt{scipy}~\cite{Virtanen:2019joe} and \texttt{gwbench}~\cite{Borhanian:2020ypi}.
\end{acknowledgments}

\putbib[bibliography]

\clearpage
\onecolumngrid
\section{End Matter}

\begin{quote}
\noindent In the End Matter, we discuss how the plunging particle gives rise to DW echoes, the limitation of the saddle point approximation, the numerical computation of the echoes, and the associated data analysis. 
\end{quote}
\twocolumngrid

\emph{\textbf{Echoes of the Plunge.---}}The radiation from the plunge is described by the $s=-2$ Teukolsky master function,
\begin{equation*}
\rho^{-4}\psi_4=\sum_{\ell m}\int d\omega\,R_{\ell m\omega}(r)\,{}_{-2}S_{\ell m\omega}(\theta)\,e^{im\phi-i\omega t},
\end{equation*}
where $\rho=-(r-ia\cos\theta)^{-1}$ and ${}_{-2}S_{\ell m\omega}$ denotes the spin-weighted spheroidal harmonics. The function $R_{\ell m \omega} \equiv R$ obeys the radial Teukolsky equation for $s=-2$ with a source $\mathcal T$ term~\cite{Xin:2021zir,SMDW}. Since $r\to\infty$, $R\to Z^\infty r^3e^{i\omega r_*}$, and since $\psi_4\to(\ddot h_+-i\ddot h_\times)/2$,
\begin{equation}
h_+-ih_\times=-\frac{2}{r}\sum_{\ell m}\int\!d\omega\,\frac{Z^\infty}{\omega^2}\,{}_{-2}S_{\ell m\omega}\,e^{im\phi-i\omega u},
\label{eq:EM_strain}
\end{equation}
where $u=t-r_*$. For a BH, the solution that is ingoing at the horizon and outgoing at infinity follows from the Green's function, built from $R^{\rm in}$ and $R^{\rm up}$.  Its amplitudes at infinity and at the horizon are~,

\begin{equation}
\begin{aligned}
    Z^\infty &=\frac{1}{2i\omega B^{\rm in}}\int dr'\,\frac{R^{\rm in}(r')\,\mathcal{T}(r')}{\Delta^2(r')}, \\
    Z^H &=\frac{B^{\rm trans}}{2i\omega B^{\rm in}C^{\rm trans}}\int dr'\,\frac{R^{\rm up}(r')\,\mathcal{T}(r')}{\Delta^2(r')},
    \label{eq:EM_Z}
\end{aligned}
\end{equation}
We sketch the derivation in Sec V of SM. Here $Z^\infty$ is the BH waveform, which contains the DW, and $Z^H$ is absorbed by the horizon. With a reflecting surface, the outgoing and ingoing amplitudes of the field near $r_*^s$ are in the ratio $\Gh=\sh\,\R\,e^{ik_H\Delta t}$. Using the above amplitudes, we can construct the ECO Green's function and the resultant ECO waveform \cite{Xin:2021zir}, 
\begin{equation}
Z^\infty_{\rm ECO}=Z^\infty+\frac{\tBH\,\Gh}{1-\rBH\,\Gh}\,Z^H=Z^\infty+\sum_{n\ge1}Z^{(n)},
\label{eq:EM_ZECO}
\end{equation}
with $Z^{(n)}=\Gh\,\tBH\,(\R\RBH e^{ik_H\Delta t})^{n-1}Z^H$. Each round trip adds the delay $\Delta t$ and a factor $\R\RBH$. Now, $n=1$ gives us the expression for the first echo, Eq.~\eqref{eq:Z1}, and its strain is given by Eq.~\eqref{eq:EM_strain} with $Z^\infty\to Z^{(1)}$, which explains the origin of the factor $1/\omega^2$ in Eq.~\eqref{eq:template}.

We can write the two homogeneous solutions to Teukolsky equation near $r_+$ as $R^{\rm in} \to B^{\rm trans}F_-$ and $R^{\rm up} \to C^{\rm up}F_++C^{\rm ref}F_-$. In Eq.~\eqref{eq:EM_Z} this splits $Z^H \simeq Z^H_{(+)}+Z^H_{(-)}$ with
\begin{equation}
\begin{aligned}
Z^H_{(+)}&=\frac{B^{\rm trans}C^{\rm up}}{2i\omega B^{\rm in}C^{\rm trans}}\int dr\,F_+\,\mathcal T\Delta^{-2},\\
Z^H_{(-)}&=\frac{B^{\rm trans}C^{\rm ref}}{2i\omega B^{\rm in}C^{\rm trans}}\int dr\,F_-\,\mathcal T\Delta^{-2}=\frac{C^{\rm ref}}{C^{\rm trans}}\,Z^\infty.
\end{aligned}
\label{eq:EM_split}
\end{equation}
Hence $\tBH Z^H_{(-)}=\rBH Z^\infty$ identically. Inserted in Eq.~\eqref{eq:Z1}, this gives the two paths of Eq.~\eqref{eq:twoparts}: the $(-)$ path is the BH waveform reflected by the barrier, $\R\RBH e^{ik_H\Delta t}Z^\infty$, while the $(+)$ path, $\R e^{ik_H\Delta t}\sh\,\tBH Z^H_{(+)}$, has no counterpart in $Z^\infty$. 
 
\begin{figure}[htbp]
\includegraphics[width=\columnwidth]{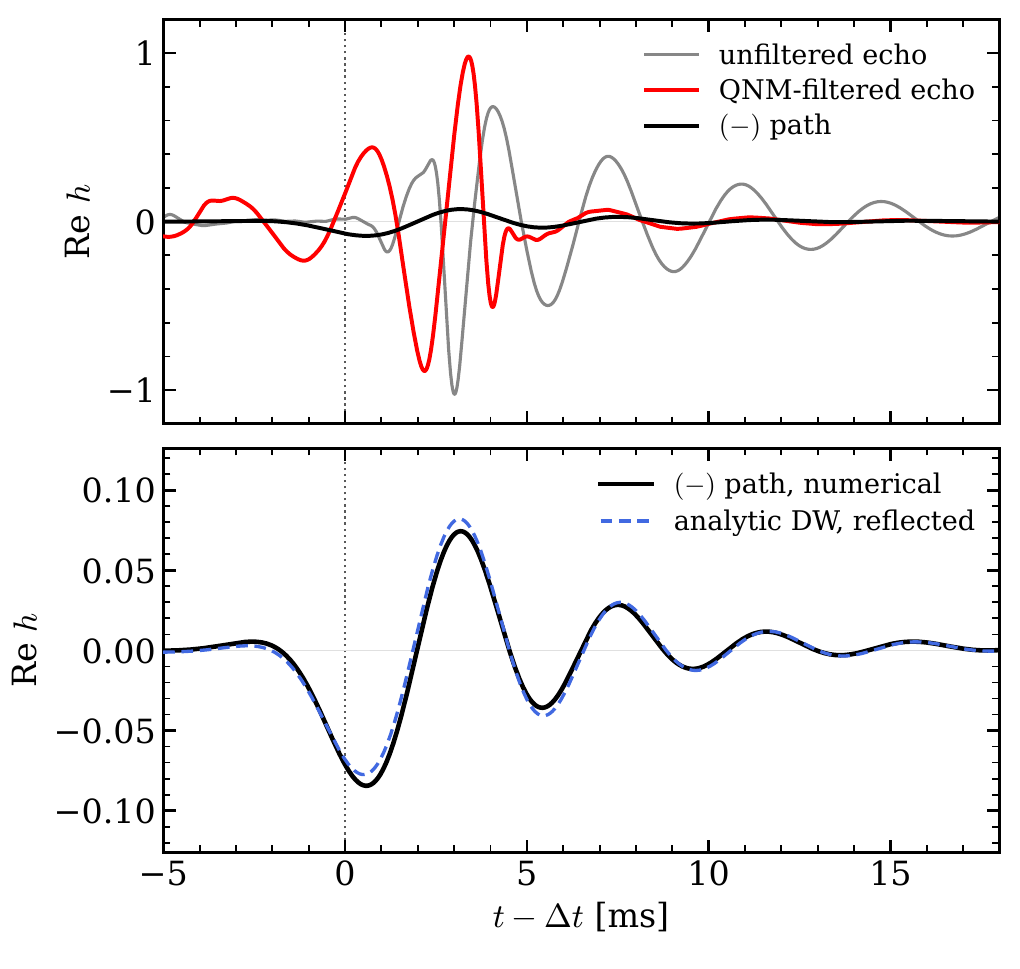}
\caption{The first echo of the plunging source, in units of its filtered peak. Top: QNM-filtered (red) and unfiltered (grey), with the $(-)$ path. The filter removes the ringing at the QNM frequencies. Bottom: $(-)$ part of numerical echo (solid) and the analytic DW reflected off the barrier with kernel $\RBH e^{i k_H \Delta t}$ (dashed).}
\label{fig:EM_echo}
\end{figure}
 
\smallskip

\emph{\textbf{Theoretical and Numerical Template.---}}To build the echo template, we first derive the analytic frequencies of the near-horizon plunge to understand the physical origins of the DW echo~(as shown in Fig.~\ref{fig:schematic}), and subsequently generate the full numerical waveform used for calibration. 

\smallskip
 
\noindent{\emph{Saddle-point analysis.}---}In Eq.~\eqref{eq:EM_split} the source is the plunging point particle. Near the horizon, where $F_-\to\Delta^2e^{-ik_Hr_*}$, the $\Delta^2$ of $F_-$ cancels the Green's-function weight $\Delta^{-2}$, and the source density of the plunge, $\propto e^{2\kappa r_*}$~\cite{Mino:2008at}, survives; the $(+)$ integral, built from $F_+\to e^{+ik_Hr_*}$, keeps $\Delta^{-2}\propto e^{-4\kappa r_*}$ (to leading order). In time domain the first echo is then $\iint dt\,d\omega\,e^{\Phi_\mp}$, with 
\begin{equation}
\Phi_\mp=-i\omega(u-\Delta t-t\pm r_*)-im(\phi\mp\Omega_Hr_*)\pm2\kappa r_*
\label{eq:EM_phase}
\end{equation}
evaluated on the trajectory, up to the constant phase $e^{-im\Omega_H\Delta t}$. The saddle-point conditions are $\partial_\omega\Phi_\mp=0$, which gives the arrival time $u=\Delta t+t\mp r_*(t)$, and, with $\dot r_*=-\beta$,
\begin{equation}
\partial_t\Phi_\mp=i\omega(1\pm\beta)-im(\dot\phi\pm\beta\Omega_H)\mp2\beta\kappa=0,
\label{eq:EM_saddle_t}
\end{equation}
which gives the frequency
\begin{equation}
\omega_\mp=\frac{m(\dot\phi\pm\beta\Omega_H)}{1\pm\beta}\mp i\,\frac{2\beta\kappa}{1\pm\beta}.
\label{eq:EM_saddle}
\end{equation}
\smallskip

\noindent\emph{The $(-)$ path.---}For $m=2$, $\omega_-=2\hat\Omega-i\hat\kappa$ is the DW frequency of Ref.~\cite{Oshita:2025qmn} [Eq.~\eqref{eq:omegaG}], and $u=t-r_*+\Delta t$ is the arrival time of the DW delayed by one round trip. The saddle is regular, $du/dt=1+\beta$: the $(-)$ path is the DW reflected by the barrier, and it reaches the observer as part of the echo.
\smallskip

\noindent\emph{The $(+)$ path.---}This is the inward emission of the plunge~\cite{Mino:2008at,Zimmerman:2011dx}, arriving at $u=v+\Delta t$ with $v=t+r_*$. As the particle approaches the horizon, $dv/dt=1-\beta\to0$, ${\rm Im}\,\omega_+=2\beta\kappa/(1-\beta)$ diverges and the integrand grows as $e^{-2\kappa r_*}$: the $(+)$ integral has no saddle point and no closed form. The decomposition serves interpretation only; for our search we use the full numerical echo described next.

\smallskip
\noindent\emph{The numerical echo.}---
Fig.~\ref{fig:EM_echo} (top) shows the first echo of the source, computed from the full $Z^H$ (as described in the Letter). Before filtering it is a short burst followed by ringing at certain frequencies, since it leaves the cavity through the same barrier as the ringdown; the Kerr QNM filter removes this ringing and leaves the non-QNM component of the echo (Fig.~\ref{fig:EM_echo}, top). Within this component the DW can be identified: reflecting the analytic DW of Eq.~\eqref{eq:dw} off the barrier, with the kernel $\RBH e^{ik_H\Delta t}$ of Eq.~\eqref{eq:twoparts}, reproduces the $(-)$ path of the numerical echo, $\RBH Z^{\infty}$ ($t > t_{0}$) with a match of $0.995$. It accounts for $10\%$ of the echo at its peak; the remainder is the $(+)$ path. To our knowledge, neither the non-QNM component of an echo nor the DW within it has been shown before. 
 
\begin{figure}[htbp]
\includegraphics[width=\columnwidth]{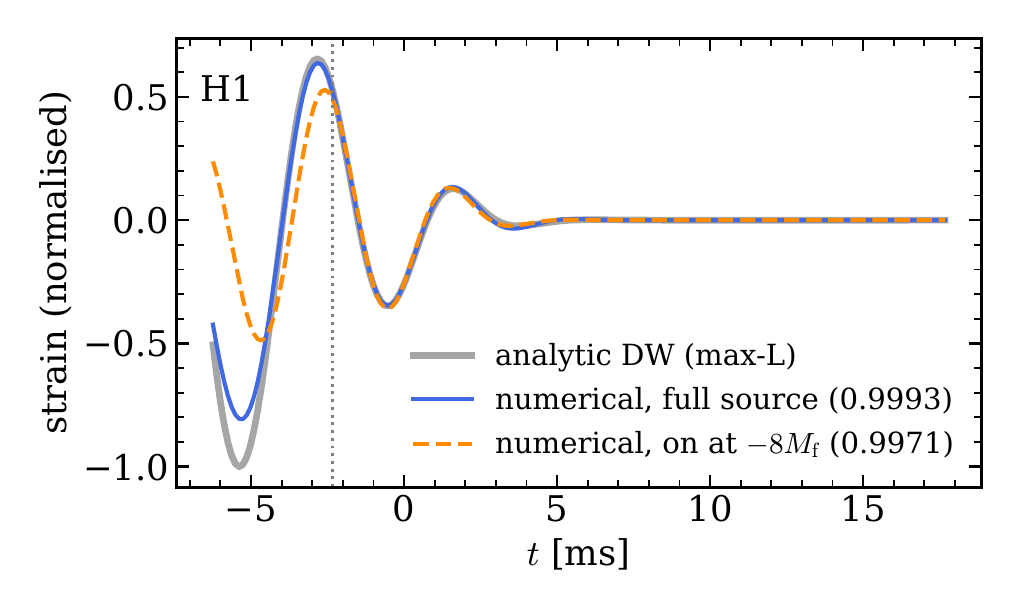}
\caption{DW of the numerical source, QNM-filtered and projected onto H1, fitted to the analytic maximum-likelihood DW (grey) for $t\geq t_0$ (right of the dotted line), with the source switched on $300\Mf$ before the light-ring crossing (blue) and at $-8\Mf$ (orange, dashed).}
\label{fig:EM_dw}
\end{figure}
 
\smallskip
\noindent{\emph{Calibration through the DW.}---}The echo is normalized by the DW (Eq. \ref{eq:template}), so the numerical source must reproduce it. The numerical DW and echo come from the same source and are filtered and projected alike. After QNM filtering and projection onto the detectors, the numerical DW matches the maximum-likelihood analytic DW with match $0.9993$ in both H1 and L1 (Fig.~\ref{fig:EM_dw}), after fit of one complex factor and one time-shift per detector. A source switched on only at $-8\Mf$ gives the same DW (match $0.997$): the DW is emitted by the final plunge. Its echo, however is larger, because the abrupt switch-on radiates towards the horizon (SM Sec.~II); the full plunge gives the smaller echo, hence the conservative bound. Despite their different onsets, the two echoes have similar shapes (fitting factor $0.85$) and differ mainly in amplitude.
 
 \smallskip
 
\textbf{\emph{Data Analysis.---}}For Gaussian noise with covariance $C_I$ and data $x_I$ in detector $I$, the likelihood of a signal $s_I$ is
\begin{equation}
\ln\mathcal L=-\tfrac12\sum_I\,(x_I-s_I)^{T}C_I^{-1}(x_I-s_I).
\label{eq:like}
\end{equation}
In the joint fit of the Letter, $s_I=\Re[a\,(d_I+\R z_I)]$, where $d_I$ is the maximum-likelihood DW, $z_I$ the QNM-filtered echo for $\R=1$ at delay $\Delta t$, and $a=A/A_{\rm ML}$ their common complex amplitude. Equation~\eqref{eq:like} is quadratic in $a$, which we integrate analytically; the DW time offset is marginalized, and the posterior and $\mathcal B$ are computed on a grid in $(\Delta t,|\R|,\phi_R)$. Holding the DW fixed, $s_I=d_I+\Re[\R z_I]$, gives the same limit (Table~\ref{tab:robust}): the two templates are nearly orthogonal, the DW peaking below $\approx250$\,Hz and the echo above. Only a prior uniform in the complex $\R$ plane, which weights large $|\R|$, raises the limit, to $0.070$--$0.073$.
 
\begin{table}[htbp!]
\caption{The 90\% upper limit on $|\R|$ and the Bayes factor with the DW held fixed and fitted together with the echo, for priors flat in $|\R|$ and in the complex plane; bold: the analysis of the Letter.}

\centering\setlength{\tabcolsep}{3pt}
\begin{tabular}{lcccc}
\hline\hline
 & \multicolumn{2}{c}{DW held fixed} & \multicolumn{2}{c}{DW fitted with the echo}\\
\cmidrule(lr){2-3}\cmidrule(lr){4-5}
prior & flat in $|\R|$ & complex $\R$ & flat in $|\R|$ & complex $\R$\\
\hline
$|\R|_{90\%}$ & $0.059$ & $0.070$ & $\mathbf{0.059}$ & $0.073$\\
$\lnB$ & $-2.9$ & $-5.7$ & $\mathbf{-2.9}$ & $-5.7$\\
\hline\hline
\end{tabular}
\label{tab:robust}
\end{table}

A negative $\lnB$ is expected even in pure noise, since the echo model has more parameters than the DW alone. GW250114 sits in the bulk of the off-source background (SM Sec.~III): its $\lnB$ is exceeded in $\sim 63\%$ of $102$ off-source segments and $61\%$ of $3000$ time slides, and its best-fitting echo (SNR $3.2$, $\Delta t\approx257$\,ms) in $\sim 73\%$ of the slides. For $400$ random injections, the credible levels of $|\R|$ and $\Delta t$ are uniform (Kolmogorov--Smirnov $p=0.10$, $0.92$), and the 90\% limit covers the injected $|\R|$ in $92.0\%$. The data show no evidence of an echo, whereas one with $|\R|=0.2$ would stand out (Fig.~\ref{fig:data}). The bound is therefore set by the absence of an echo, not by a lack of sensitivity. 

\end{bibunit}

\clearpage
\onecolumngrid

\makeatletter
\def\set@footnotewidth{\onecolumngrid}
\def\footnoterule{\kern-6pt\hrule width 1.5in\kern6pt}
\makeatother

\pdfbookmark[0]{Supplemental Material}{supp}

\vspace*{0.5cm}
\noindent\rule{\linewidth}{0.5pt}
\vspace{0.8cm} 

\begin{center}
\textbf{\Large Supplemental Material for:\\[4pt] Listening to the Horizon: Probing Near-Horizon Reflectivity with GW250114}
\end{center}

\vspace{0.4cm} 

\begin{center}
R. Prasad~\orcidlink{0000-0002-6602-3913}$^{1,\,\ast}$\ \ 
Subhodeep Sarkar~\orcidlink{0000-0001-9015-8837}$^{1,\,\dagger}$\ \ and\ \ 
Sudipta Sarkar~\orcidlink{0000-0002-4077-1358}$^{1,\,\ddagger}$\\[0.3cm]
\textit{$^{1}$Department of Physics, Indian Institute of Technology Gandhinagar, Palaj, Gandhinagar 382055, Gujarat, India}
\end{center}

\vspace{0.8cm} 

\renewcommand{\thepage}{SM--\arabic{page}}
\setcounter{page}{1}

\renewcommand{\thefigure}{S\arabic{figure}}
\setcounter{figure}{0}

\makeatletter
\renewcommand{\theHfigure}{S.\arabic{figure}}
\renewcommand{\theHtable}{S.\arabic{table}}
\renewcommand{\theHequation}{S.\arabic{equation}}
\makeatother

\renewcommand{\theequation}{S\arabic{equation}}
\setcounter{equation}{0}

\renewcommand{\thetable}{S\arabic{table}}
\setcounter{table}{0}

\begin{quote}
\small 
\noindent The following pages contain supplemental material accompanying the main text. This includes additional derivations, reproduction of known results, code validation, analysis details, and figures that support the results and conclusions presented in the Letter. Readers seeking a self-contained summary of the results should refer to the main text; the material here is intended to be read alongside it.
\end{quote}

\vspace{0.5cm} 

\vfill

\noindent\rule{0.3\linewidth}{0.4pt}\\[3pt]
{\footnotesize
$\ast$ \href{mailto:prasad@iitgn.ac.in}{prasad@iitgn.ac.in}\\
$\dagger$ \href{mailto:ssarkar@iitgn.ac.in}{ssarkar@iitgn.ac.in}\\
$\ddagger$ \href{mailto:sudiptas@iitgn.ac.in}{sudiptas@iitgn.ac.in}
}

\clearpage

\begin{bibunit}
\makeatletter\global\let\@FMN@list\@empty\makeatother

\providecommand{\upd}[1]{\textcolor{red}{#1}}
 
\suppsection{Data, QNM filter and the direct wave}
 
\emph{\textbf{Data.---}} We use the H1 and L1 strain of GW250114~\cite{GW250114_release_sup, LIGOScientific:2025rid_sup}. As in Ref.~\cite{Lu:2025vol_sup}, the noise covariance of each detector is computed from the autocorrelation of $64$\,s of data starting $1$\,s after the strain peak. The remnant is fixed to the maximum-posterior values of Ref.~\cite{LIGOScientific:2025snk_sup}, $\Mf=67.93\,M_\odot$ (detector frame) and $\chi_f=0.6725$, for which $m\Omega_H/2\pi=184$\,Hz and $\kappa^{-1}=1.57$\,ms.
 
\emph{\textbf{QNM filter.}---} With the convention $h\propto e^{-i\omega t}$, the rational filter~\cite{Ma:2022wpv_sup}
\begin{equation}
F(\omega)=\prod_{n=0}^{2}\frac{\omega-\omega_{22n}}{\omega-\omega_{22n}^*}\,\frac{\omega+\omega_{22n}^*}{\omega+\omega_{22n}}
\label{eq:S_filter}
\end{equation}
cancels the $(2,2,n\le2)$ QNM poles of a real strain and is all-pass, $|F(\omega)|=1$ for real $\omega$. We apply it to the data and to the numerical templates of Sec.~{II}. The analytic DW contains no QNMs and is fitted to the filtered data without filtering, as in Ref.~\cite{Lu:2025vol_sup}.
 
\emph{\textbf{Direct Wave model.---}} The DW template is given by Eqs.~\eqref{eq:dw} and \eqref{eq:omegaG} of the Letter, with $\omega_H=2\Omega_H-i\kappa$, evaluated along the prograde equatorial geodesic that plunges from the ISCO of the remnant with the ISCO energy $E$ and angular momentum $L$ \cite{Oshita:2025qmn_sup},
\begin{equation}
r^2\frac{dr}{d\tau}=-\sqrt{P^2-\Delta\big[r^2+(L-aE)^2\big]},\qquad
r^2\frac{dt}{d\tau}=\frac{(r^2+a^2)P}{\Delta}-a(aE-L),\qquad
r^2\frac{d\phi}{d\tau}=\frac{aP}{\Delta}-(aE-L),
\label{eq:S_geo}
\end{equation}
where $P=E(r^2+a^2)-aL$. The time $t$ in Eq.~\eqref{eq:dw} is the retarded time of emission, $t_e-r_*(t_e)$, at which $\omega_G$ is evaluated. The prefactor $\omega_G-\omega_H$ vanishes at the horizon-mode frequency, so that $\omega_H$, which is screened by the potential barrier, does not contribute~\cite{Oshita:2025qmn_sup}. The normalization, a constant phase and the origin of $t$ are absorbed into $A$ and $\delta t$.
 
\begin{figure}[!htb]
\centering
\includegraphics[width=0.42\textwidth]{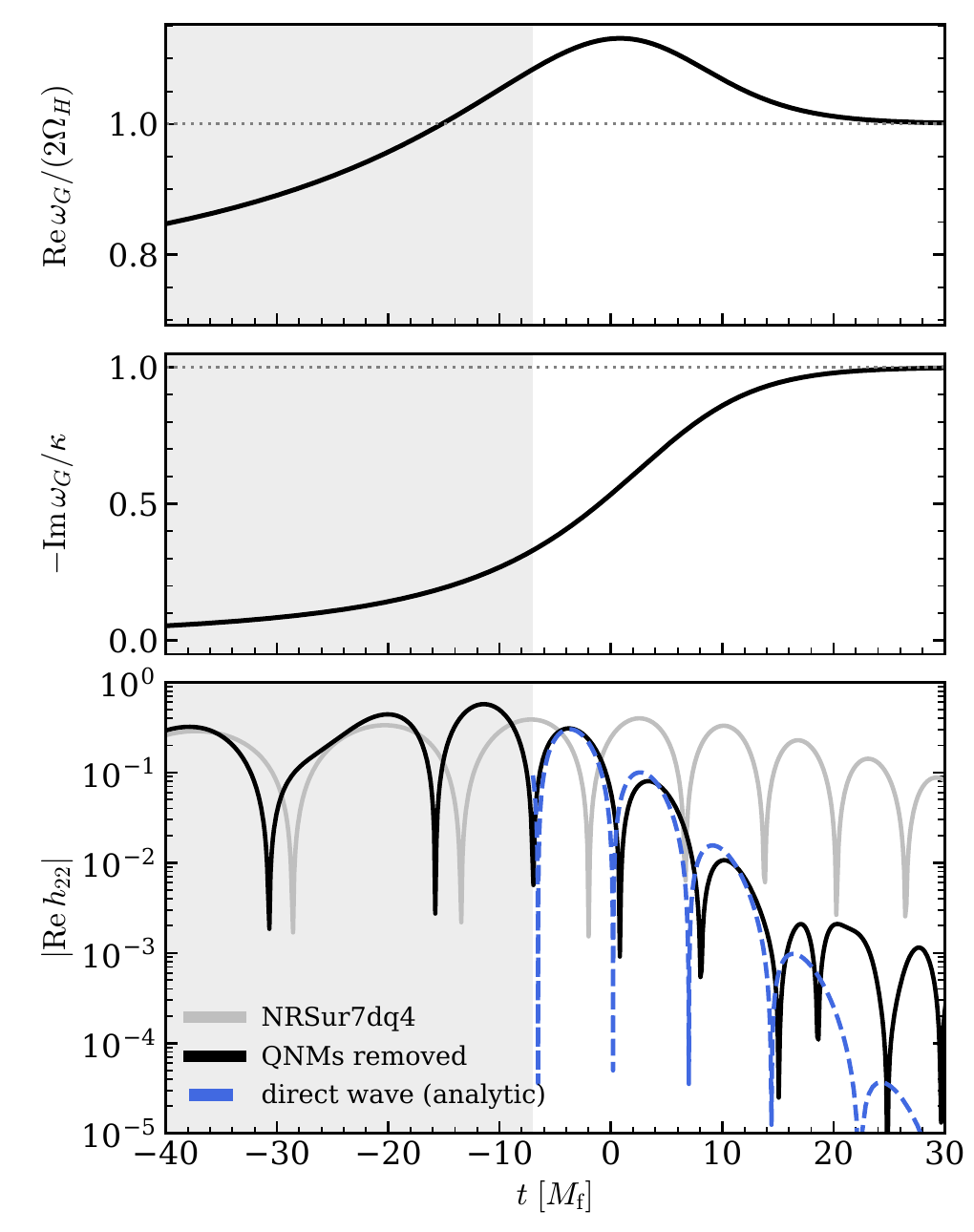}\hfill
\includegraphics[width=0.55\textwidth]{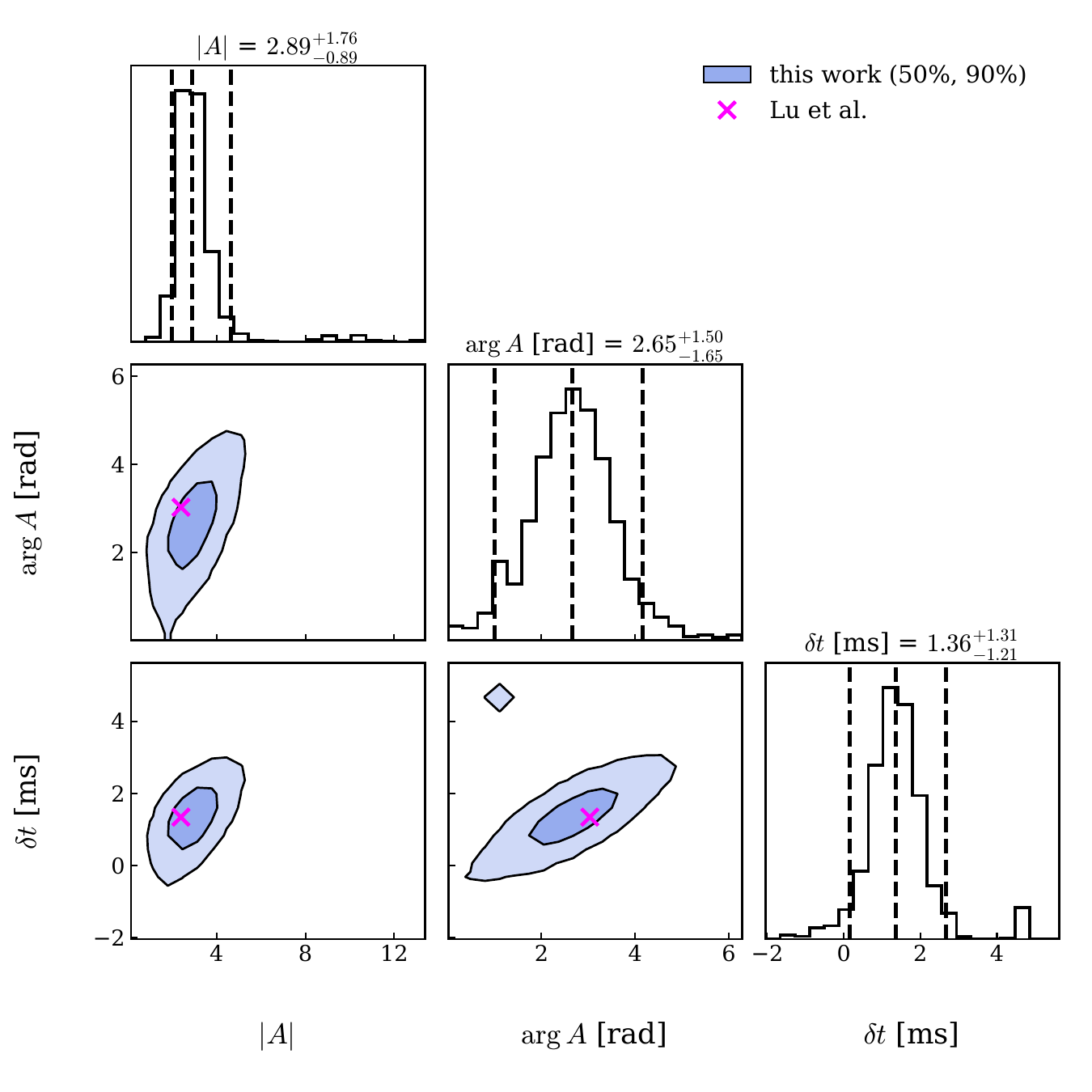}
\caption{The direct wave. Left: $\mathrm{Re}\,\omega_G$ (top) and $-\mathrm{Im}\,\omega_G$ (middle) along the plunge, in units of $2\Omega_H$ and $\kappa$; bottom: the $(2,2)$ mode of the NRSur7dq4 waveform of GW250114 before (grey) and after (black) QNM removal, with the analytic DW (dashed). Time is in units of $\Mf$ from the strain peak; shaded: $t<t_0$, not fitted part. Right: posterior of the DW amplitude $|A|$, phase $\arg A$ and time offset $\delta t$, with 50\% and 90\% credible regions; titles give the medians and 90\% intervals. Crosses are estimates of Ref.~\cite{Lu:2025vol_sup}.}
\label{fig:S1}
\end{figure}

Along the plunge, $\mathrm{Re}\,\omega_G$ approaches $2\Omega_H$ and $-\mathrm{Im}\,\omega_G$ rises toward $\kappa$ (Fig.~\ref{fig:S1}, left), so the DW is neither a QNM nor a horizon mode of fixed frequency. From $t_0=-7\Mf$ on, with its amplitude and time offset fitted, it follows the QNM-filtered $(2,2)$ mode of the maximum-posterior NRSur7dq4 waveform of GW250114 (Fig.~\ref{fig:S1}, bottom left).

In detector $I$ the template is
\begin{equation}
h_I=F_{+,I}\,h_++F_{\times,I}\,h_\times,\qquad
h_+-ih_\times=A\,h_{\rm DW}(t-\delta t)\,{}_{-2}Y_{22}(\iota,\varphi)+A^*h^*_{\rm DW}(t-\delta t)\,{}_{-2}Y_{2,-2}(\iota,\varphi),
\label{eq:S_proj}
\end{equation}
with the antenna patterns, arrival times, inclination and polarization angle of the maximum-posterior sample of GW250114.
 
\smallskip
\emph{\textbf{Fit to GW250114.---}} We fit $A=A_x+iA_y$ and $\delta t$ to the QNM-filtered data on the $0.32$\,s segment that starts at $t_0=-7\Mf$ ($-2.3$\,ms) from the strain peak, with the likelihood of Eq.~\eqref{eq:like} (End Matter), priors uniform in $A_x,A_y\in[-20,20]$ and $\delta t\in[-20,20]\,\Mf$, and the ensemble sampler \texttt{emcee}. The posterior (Fig.~\ref{fig:S1}, right) is consistent with that of Ref.~\cite{Lu:2025vol_sup}: the medians of $(|A|,\arg A,\delta t)$ are $(2.89,\,2.65\,{\rm rad},\,1.36\,{\rm ms})$ against $(2.86,\,3.05\,{\rm rad},\,1.34\,{\rm ms})$. The maximum-likelihood DW has matched-filter SNRs of $16.0$ (H1) and $17.1$ (L1), against $15.8^{+0.1}_{-0.5}$ and $17.1^{+0.1}_{-0.4}$ in Ref.~\cite{Lu:2025vol_sup}. Its amplitude and phase normalize the echo template (Sec.~II).
\smallskip
 
\emph{\textbf{Echo delay and surface position.---}} By Eq.~\eqref{eq:dt} of the Letter, the delay grows by $4r_+\Mf/(r_+-r_-)=\kappa^{-1}$. Since $\epsilon=r_s-r_+$ is a Boyer--Lindquist coordinate offset, the physical distance of the surface from the horizon is the proper radial distance, $d_s\simeq2r_+[\epsilon/(r_+-r_-)]^{1/2}$ at the equator. A surface one Planck length from the horizon, $d_s=\ell_{\rm P}$, as in Refs.~\cite{Abedi:2016hgu_sup,Cardoso:2019rvt_sup}, has $\epsilon\simeq4\times10^{-81}\Mf$ (for the source-frame remnant mass, $z=0.085$) and $\Delta t\simeq 292$\,ms. The searched range, $\Delta t=12$--$300$\,ms, covers $10^{-83}\lesssim\epsilon/\Mf\lesssim10^{-3}$ ($10^{-41}\lesssim d_s/\Mf\lesssim10^{-1}$) and includes it; its lower end keeps the echo well separated from the DW.
 
\suppsection{Numerical Teukolsky echoes}
 
\emph{\textbf{Method.---}} To obtain the numerical echoes we follow the approach described in Ref.~\cite{Xin:2021zir_sup}. A point particle moves on an adiabatic sequence of circular equatorial orbits whose orbital frequency is half the $(2,2)$ frequency of an equal-mass ($q=1$) waveform, matched in retarded time; we use the NR-calibrated SEOBNRv4 model in place of the NRSur7dq4 waveform of Ref.~\cite{Xin:2021zir_sup}. At the ISCO the radiation reaction is switched off and the particle follows the plunging geodesic of Eq.~\eqref{eq:S_geo}, with $E$ and $L$ fixed by its four-velocity there: the ISCO angular velocity and the radial velocity reached by the inspiral. The late inspiral, and with it the initial conditions of the plunge, therefore follow the comparable-mass waveform, while the plunge itself is geodesic. The source is switched on smoothly $300\Mf$ before the particle crosses the light ring, so that the whole plunge contributes. 

We compute $Z^\infty$ and $Z^H$ with \texttt{pybhpt} for $0.02\le\Mf\omega\le4$ ($f\lesssim1.9$\,kHz) and continue the radial functions to the horizon with a Frobenius series. The barrier amplitudes satisfy the flux identity $|\RBH|^2+\Gamma=1$, with $\Gamma$ computed from $\tBH$, to $6\times10^{-4}$ between $30$ and $800$\,Hz.

\smallskip

\begin{figure}[tbp]
\centering
\includegraphics[width=0.9\textwidth]{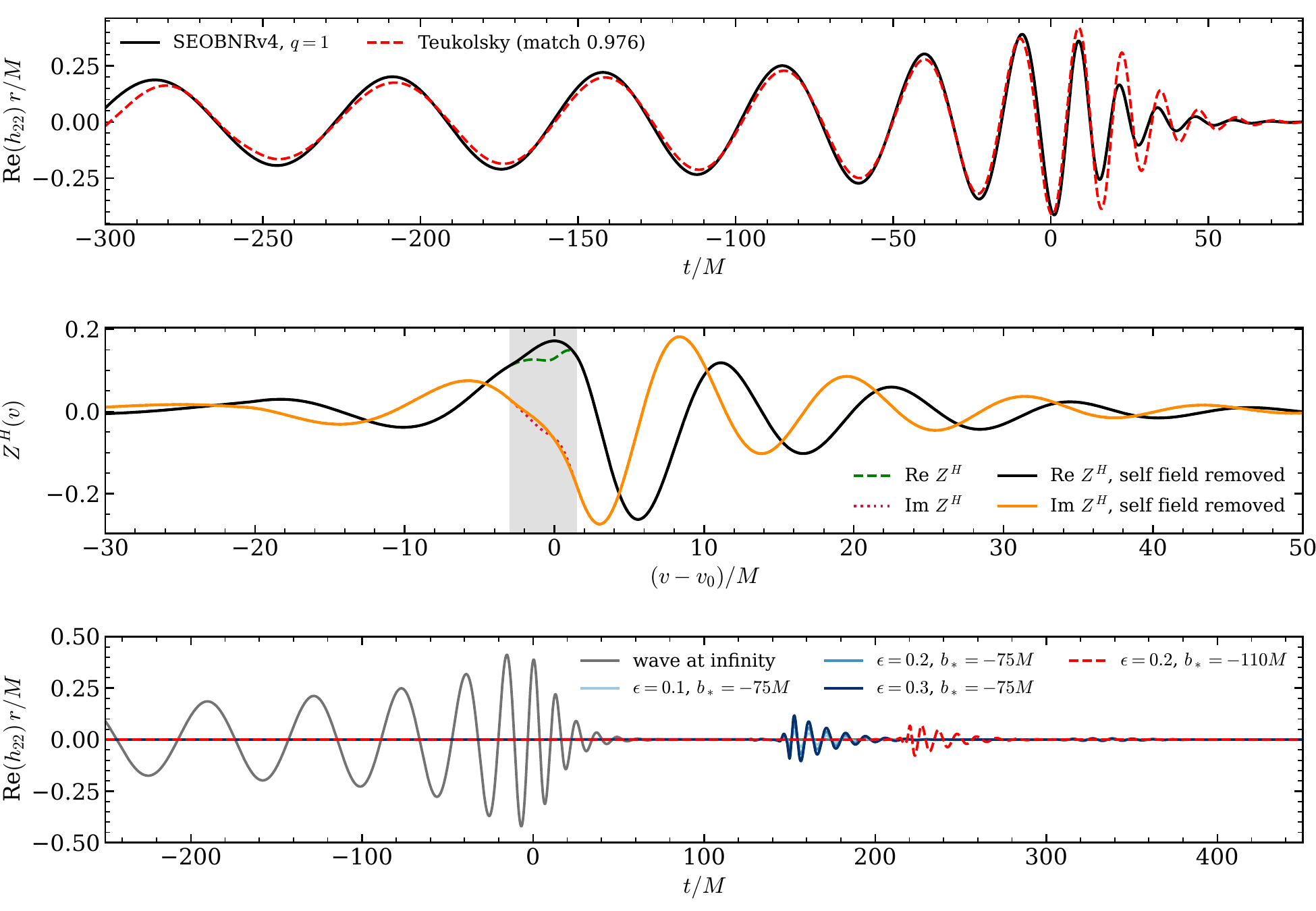}
\caption{Validation of the numerical source in the setup of Ref.~\cite{Xin:2021zir_sup}. Top: the $(2,2)$ Teukolsky waveform against SEOBNRv4 for $q=1$ (match $0.976$). Middle: the horizon amplitude $Z^H$ around the horizon crossing of the particle, at advanced time $v_0$, before (dashed, dotted) and after (solid) removal of the self-field feature; shaded: the replaced interval. Bottom: the wave at infinity and its echoes for Lorentzian reflectivities of width $M\Gamma=0.5$, labeled by their amplitude $\epsilon$ and surface position $b_*=r_*^s$ (notation of Ref.~\cite{Xin:2021zir_sup}).}
\label{fig:S2}
\end{figure}

\emph{\textbf{Validation.---}} For the setup of Ref.~\cite{Xin:2021zir_sup}, our Teukolsky waveform agrees with SEOBNRv4 (match $0.976$), and our echoes for their Lorentzian reflectivities agree with their result (echo-to-wave peak ratios); Fig.~\ref{fig:S2}. As in Ref.~\cite{Xin:2021zir_sup}, we remove the non-radiative self-field feature of $Z^H$ where the particle crosses the horizon, at advanced time $v_0$, by replacing $Z^H(v)$ on $v_0-3\Mf<v<v_0+1.5\Mf$ with a cubic spline through the neighboring samples.

\smallskip
\emph{\textbf{Numerical and analytic DW.---}} The QNM-filtered direct wave of the numerical source and the analytic DW of Sec.~I describe the same emission. After projection onto H1 and L1 and a fit of one complex amplitude and one time shift, the numerical DW matches the maximum-likelihood analytic DW on the analysis segment with match $0.9993$ (End Matter, Fig.~\ref{fig:EM_dw}). A source switched on only at $-8\Mf$, close to $t_0$, gives the same DW (match $0.9971$): the DW is emitted by the final plunge. The amplitude and shift of the full source normalize its echo.

\begin{figure}[!htb]
\centering
\begin{minipage}[c]{0.5\textwidth}\centering
\includegraphics[width=0.9\linewidth]{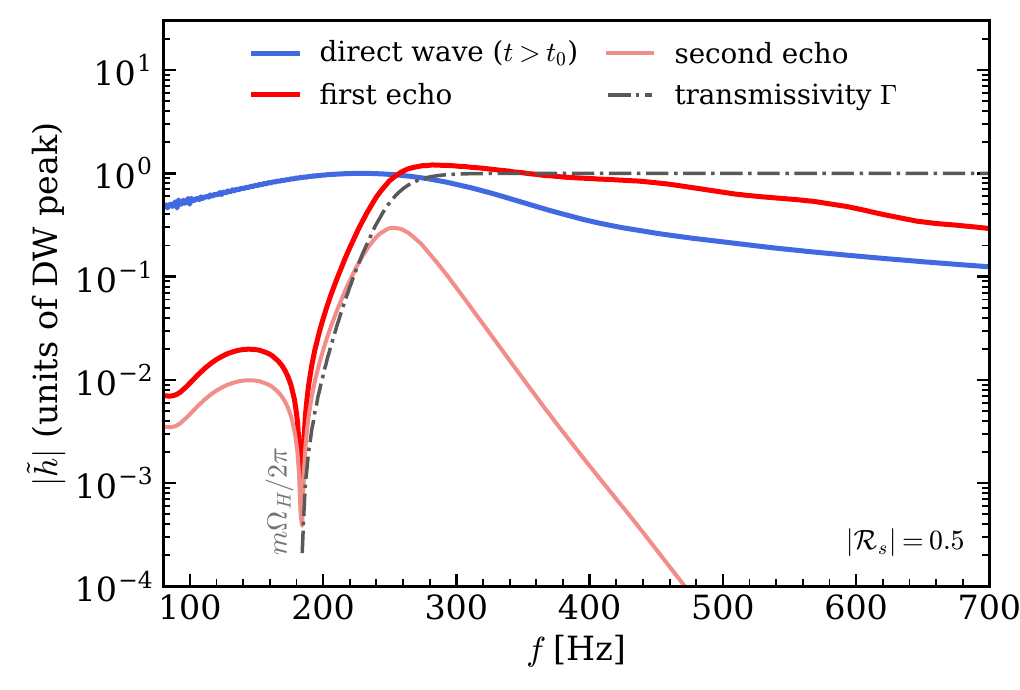}
\end{minipage}\hspace{0.03\textwidth}
\begin{minipage}[c]{0.44\textwidth}
\caption{Spectra of the DW at $t>t_0$ (blue) and of the first and second echoes (red, light red) of Fig.~\ref{fig:template} ($|\R|=0.5$, $\Delta t=25$\,ms), in units of the maximum of the DW spectrum, with the barrier transmissivity $\Gamma=1-|\RBH|^2$ (dash-dotted). The echo vanishes at $f=m\Omega_H/2\pi$, where $\Gamma=0$, and rises with $\Gamma$ above it.}
\label{fig:S_spec}
\end{minipage}
\end{figure}

\smallskip
\emph{\textbf{The echo.---}} Before filtering, the first echo is a short burst followed by ringing at the QNM frequencies of the remnant, dominated by the $(2,2,0)$ mode ($f_{220}\simeq250$\,Hz): it leaves the cavity through the same potential barrier as the ringdown. The QNM filter, applied to the data and the template alike, removes the ringing and leaves the non-QNM component of the echo (End Matter, Fig.~\ref{fig:EM_echo}, top). This component has two parts (Fig.~\ref{fig:EM_echo}, bottom). The $(-)$ path, the DW reflected inward by the barrier, $\RBH Z^\infty$ (DW $t > t_0$), is reproduced by reflecting the analytic DW off the barrier (match $0.995$, amplitude within $4\%$); it is the inside prescription of Refs.~\cite{Wang:2019rcf_sup,Maggio:2019zyn_sup} applied to the DW, and only $10\%$ of the echo at its peak. The rest is the $(+)$ path, radiation emitted toward the horizon during the plunge, which has no counterpart in the DW. In frequency (Fig.~\ref{fig:S_spec}), the echo vanishes at the superradiant threshold $f=m\Omega_H/2\pi=184$\,Hz, where $\Gamma=0$, and follows $\Gamma$ above it, so that most of its power lies above $\approx250$\,Hz, while the DW lies below. The second echo carries a further factor $\R\RBH$ and is negligible.

\smallskip
 
\begin{figure}[!htb]
\centering\includegraphics[width=0.9\textwidth]{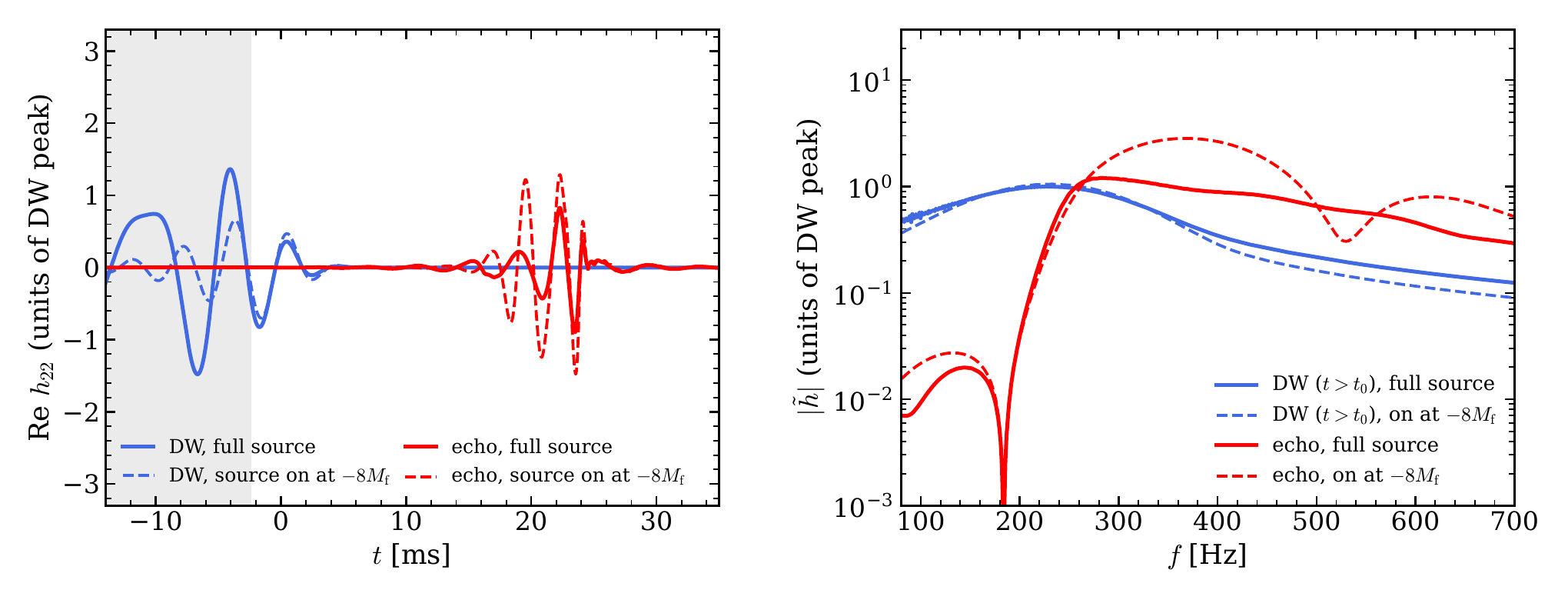}
\caption{Same DW, different echoes. Left: DW and first echo ($|\R|=0.5$, $\Delta t=20$\,ms) for the full source (solid) and the source switched on at $-8\Mf$ (dashed). Right: their spectra (echo for $|\R|=0.5$).}
\label{fig:S6}
\end{figure}

\textbf{\emph{Template systematics.---}} Unlike the DW, the echo depends on how the source is modeled near the horizon. A source switched on at $-8\Mf$ reproduces the DW equally well (match 0.997), but its abrupt switch-on radiates toward the horizon and makes the echo comparatively larger, mostly at $300$--$450$\,Hz (Fig.~\ref{fig:S6}); keeping the self field also enlarges it. Injected into noise and recovered with the baseline template, these variants give fitting factors of $0.85$ and $0.97$ and recovered $|\R|$ larger than injected by factors $2.6$ and $1.14$ (for $R_s = 0.1$). The baseline (full inspiral--plunge source, self field removed) therefore gives the smallest echo of these choices, and hence the most conservative bound; either variant (late switch-on or self-field) would tighten it.

\smallskip
\emph{\textbf{Match.---}} To compare two waveforms we use their match~\cite{Owen:1995tm_sup},
\begin{equation}
\mathcal M(a,b)=\max_{\tau}\frac{|(a_\tau,b)|}{\sqrt{(a_\tau,a_\tau)\,(b,b)}},\qquad (x,y)=\sum_{t_k\in W}x^*(t_k)\,y(t_k),
\label{eq:S_match}
\end{equation}
with the sum over the samples in a time window $W$ ($[t_0,t_0+47\Mf]$ for the DW). The match uses no PSD: all frequencies are weighted equally, so it compares the shapes of the waveforms alone. The noise-weighted quantities (likelihood, SNRs, fitting factors~\cite{Apostolatos:1995pj_sup} and the DW--echo overlap) use the noise covariance $C_I$ of each detector, computed from the autocorrelation of $64$\,s of data starting $1$\,s after the strain peak (Sec.~I); the forecast uses the design noise curves of Sec.~IV.

\suppsection{The search on GW250114 and tests of the bound}
 
\emph{\textbf{Likelihood.---}} All analyses use the Gaussian likelihood of Eq.~\eqref{eq:like} of the End Matter; they differ only in the signal model $s_I$. The echo is linear in $\R$, so at each delay the likelihood is an exact Gaussian in the complex $\R$ and is evaluated on a grid rather than sampled. We analyze the QNM-filtered H1 and L1 data on the $0.32$\,s segment that starts at $t_0$. The template $z_I(t;\Delta t)$ is the first echo of Sec.~II for $\R=1$, projected onto detector $I$ as the DW in Eq.~\eqref{eq:S_proj}; a change of $\Delta t$ delays it and multiplies it by a constant phase, which the uniform prior on $\phi_R$ absorbs. With the DW held fixed at its maximum-likelihood value, $s_I=d_I+\Re[\R z_I]$, and Eq.~\eqref{eq:like} becomes, in terms of the residual $r_I=x_I-d_I$,
\begin{equation}
\ln\frac{\mathcal L}{\mathcal L_0}=b\cdot x-\tfrac12\,x^{T}Fx,\qquad x=(\Re\R,\Im\R),
\label{eq:S_quad}
\end{equation}
with $b_a=\sum_I\langle S_{a,I}|r_I\rangle$, $F_{ab}=\sum_I\langle S_{a,I}|S_{b,I}\rangle$, $S_{1,I}=\Re z_I$, $S_{2,I}=-\Im z_I$ and $\langle u|v\rangle=u^TC_I^{-1}v$. The echo SNR per unit $|\R|$ is $\rho_1=\sqrt{{\rm tr}F/2}$, at all delays. The grid has $400\times144$ points in $(|\R|,\phi_R)$ and $2881$ delays spaced by $0.1$\,ms. The spacing must resolve the $\approx1$\,ms echo burst. The prior is uniform in $|\R|\le1$, $\phi_R$ and $\Delta t$; we also quote a prior uniform in the complex plane, $p(|\R|)\propto|\R|$. The Bayes factor $\mathcal B$ of echo against no echo is the prior average of $\mathcal L/\mathcal L_0$. Evaluating the likelihood with the filter and noise covariance applied at every point, or from filtered and whitened templates stored once, gives the same $\lnB=-2.88$, and limits within $0.002$; the stored templates are used for the background and the injections.
 
\smallskip
\emph{\textbf{Constraints.---}} With the DW held fixed, the echo alone gives $|\R|<0.059$ and $\lnB=-2.9$ for the prior uniform in $|\R|$, and $|\R|<0.070$ and $\lnB=-5.7$ for the prior uniform in the plane. At fixed $\Delta t$ the limit ranges from $0.029$ to $0.077$, with no trend with the position of the surface.
\smallskip 
 
\begin{figure}[!htb]
\centering
\begin{minipage}[c]{0.52\textwidth}
\centering
\includegraphics[scale=0.5]{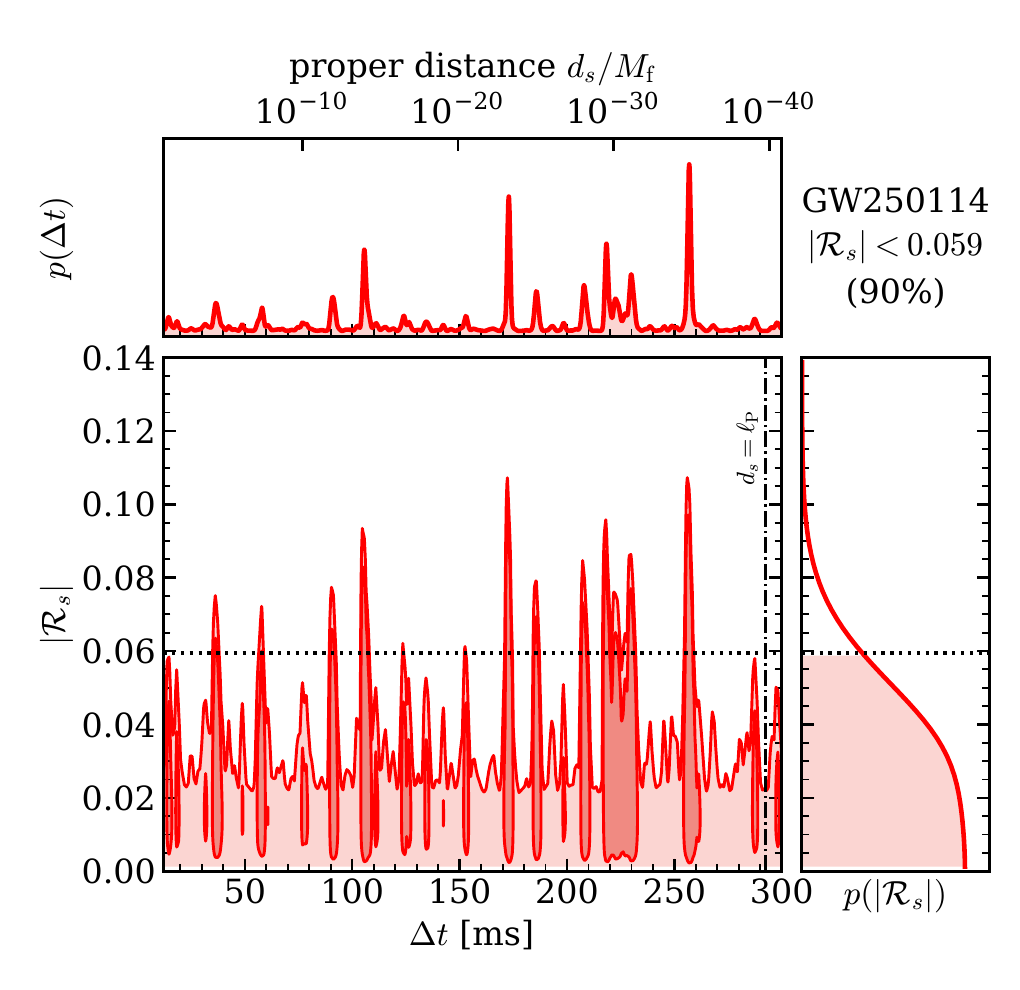}
\end{minipage}\hfill
\begin{minipage}[c]{0.45\textwidth}
\caption{Joint posterior of $\Delta t$ and $|\R|$ for GW250114 (50\% and 90\% credible regions), with the marginals of $\Delta t$ (top; top axis: proper distance $d_s/\Mf$) and $|\R|$ (right). Dotted: the 90\% limit; dash-dotted: a surface one Planck length (proper distance) from the horizon.}
\label{fig:S7}
\end{minipage}
\end{figure}

\begin{figure}[!htb]
\centering\includegraphics[width=0.85\textwidth]{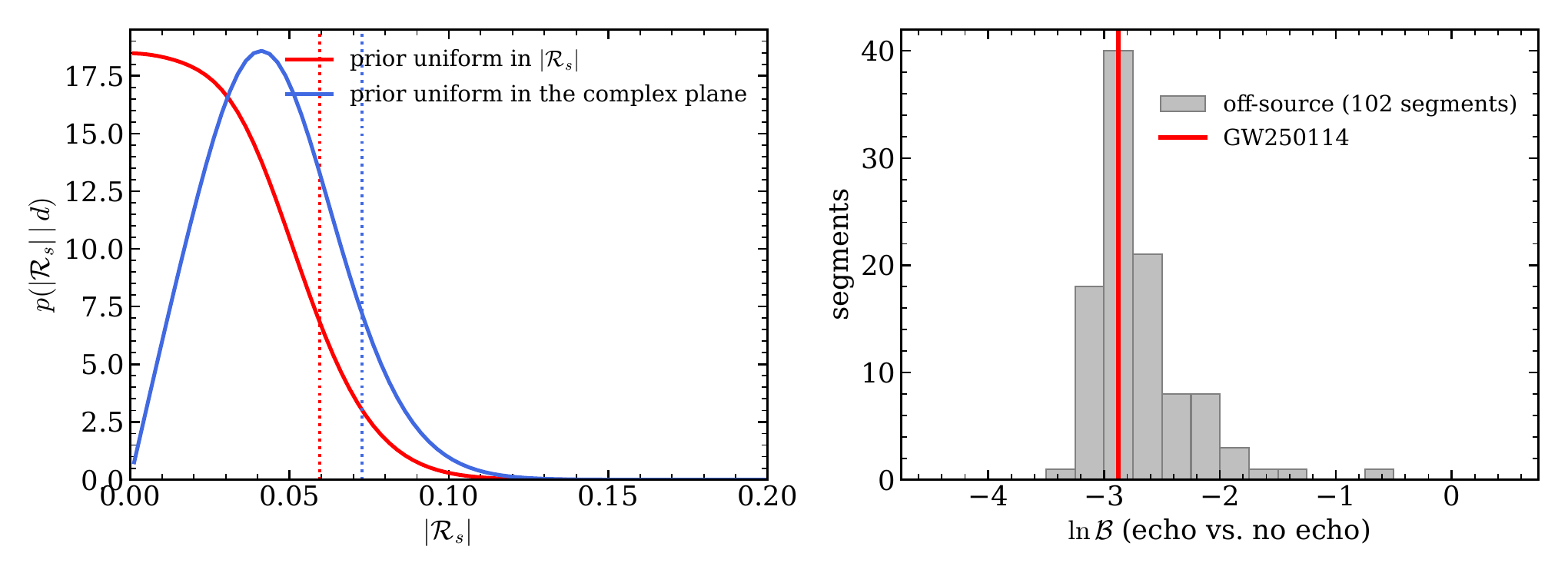}
\caption{Left: posterior of $|\R|$ for GW250114 from the joint fit of the DW and the echo, for priors uniform in $|\R|$ (red) and in the complex $\R$ plane (blue); dotted: the 90\% limits, $0.059$ and $0.073$. Right: $\lnB$ of the echo model for GW250114 (red) and for $102$ off-source segments with the maximum-likelihood DW added, analyzed in the same way (grey).}
\label{fig:S_bg}
\end{figure}

In the analysis of the Letter, the DW and the echo are fitted together, $s_I=\Re[a\,(d_I+\R z_I)]$, with $d_I$ the maximum-likelihood DW and $a$ a free complex factor common to both. Here $a=A/A_{\rm ML}$ is the complex amplitude $A$ of Eq.~\eqref{eq:template} in units of its maximum-likelihood value. At fixed $(\R,\Delta t)$ the model is linear in $a$, and integrating Eq.~\eqref{eq:like} over $a$ with a flat prior gives, up to a constant common to the models with and without echo, the marginal likelihood
\begin{equation}
\ln\mathcal L(\R,\Delta t,\delta t)=\tfrac12\,b_G^{T}F_G^{-1}b_G-\tfrac12\ln\det F_G,
\label{eq:S_jointfit}
\end{equation}
where $b_G$ and $F_G$ are the projections of the data on the two quadratures of $d+\R z$ and their Gram matrix. The common time offset $\delta t$ is marginalized within $\pm1$\,ms of its best value, and $\mathcal B$ is taken against the DW alone. The joint fit gives $|\R|<0.059$ and $\lnB=-2.9$, and $|\R|<0.073$ and $\lnB=-5.7$ for the prior uniform in the plane (Figs.~\ref{fig:S7} and \ref{fig:S_bg}, left; Table~\ref{tab:robust}). It agrees with the analysis with the DW held fixed because the two templates are nearly orthogonal: the DW lies below $\approx250$\,Hz and the echo above, and their normalized overlap is at most $0.05$. A Markov-chain Monte Carlo over all six parameters, which frees the DW completely, gives $|\R|<0.070$ for the prior uniform in the plane, as the grid, and does not change the bound.

\smallskip 
\emph{\textbf{Comparison with noise.---}} The posterior of $\Delta t$ is broad (Fig.~\ref{fig:S7}), with a highest peak at $\Delta t\approx 257$\,ms ($\Delta\ln\mathcal L=5.2$, echo SNR $3.2$), and $\lnB$ is negative. Neither number can be judged on its own: a model with more parameters than the null gives a negative $\lnB$ in pure noise, and some delay always fits best. We therefore repeat the search on $102$ non-overlapping off-source segments recorded $16.6$--$49.3$\,s after the event, each with the maximum-likelihood DW added and analyzed as the on-source segment but with the DW time offset fixed at its maximum-likelihood value (the on-source result is compared at the same fixed offset), and on $3000$ time slides, analyzed with the DW held fixed, that pair the H1 data of one segment with the L1 data of another, which removes any coherent signal while keeping the noise of each detector. GW250114 is typical of this background (Fig.~\ref{fig:S_bg}, right): its $\lnB$ is exceeded in $63\%$ of the segments and $61\%$ of the time slides, and its highest peak in $73\%$ of the slides. The 90\% limit in noise has median $0.063$ and 90\% range $0.052$--$0.087$, and in $6$\,ms bins of $\Delta t$ the on-source limit lies within the corresponding range in all but a few of the bins, consistent with chance [Fig.~\ref{fig:constraint} (top)].
 
\smallskip
 
\emph{\textbf{Would the search have found an echo?---}} A bound is meaningful only if an echo above it would have been detected and if the posterior is calibrated. We test both by injecting echoes of the search template into the off-source segments and repeating the full search: echoes at fixed $|\R|=0$--$0.3$ and $\Delta t=20$ and $30$\,ms ($50$ segments each), and $400$ echoes with $|\R|$, $\Delta t$ and $\phi_R$ drawn uniformly from $[0,0.3]$, $[12,300]$\,ms and $[0,2\pi)$, at arbitrary delays rather than on the grid. Both sets are repeated with the DW injected and fitted together with the echo.

\begin{figure}[!htb]
\centering\includegraphics[scale=0.48]{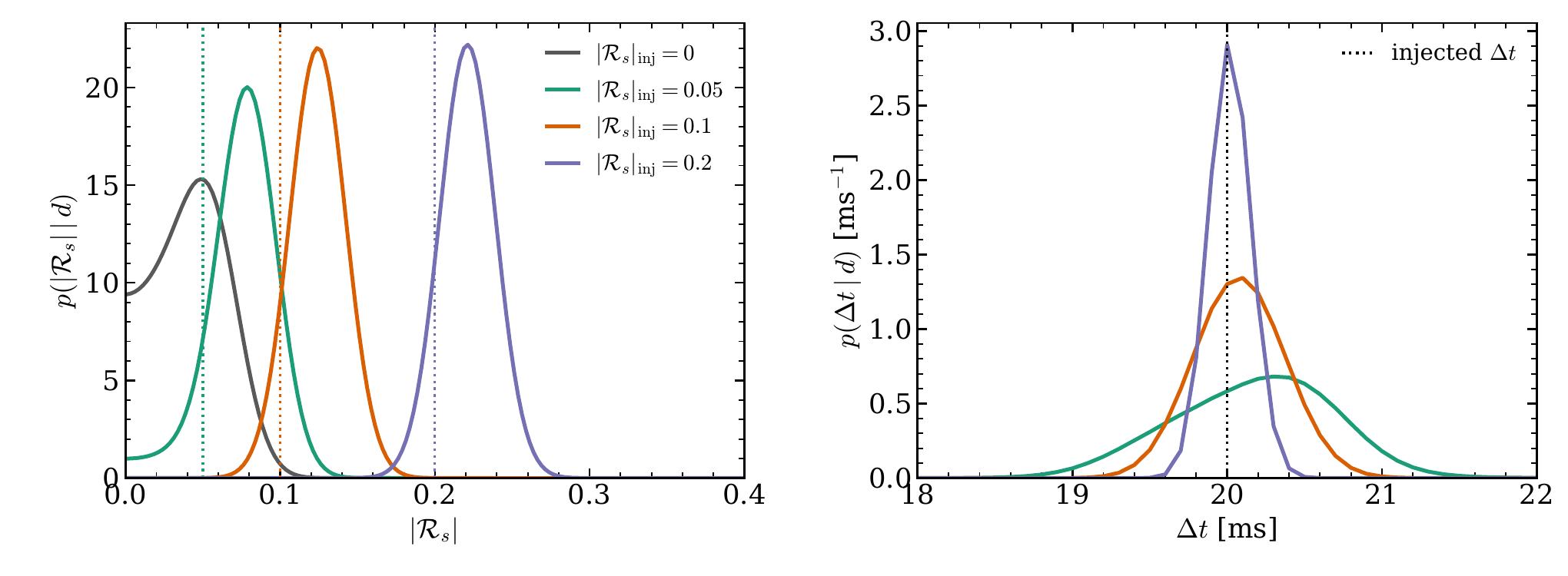}
\caption{Echoes of $|\R|=0$, $0.05$, $0.1$ and $0.2$ ($\Delta t=20$\,ms) injected into one off-source segment. Left: posterior of $|\R|$. Right: posterior of $\Delta t$. Dotted: injected values.}
\label{fig:S10}
\end{figure}

An echo near the bound, $|\R|=0.05$ (SNR $\approx3$), shifts the posterior in the segment of Fig.~\ref{fig:S10}, but over $50$ segments its delay is recovered within $0.5$\,ms in only $16$--$26\%$ of cases, whereas at $|\R|=0.1$ (SNR $\approx6$) this holds in $92$--$96\%$, and both $|\R|$ and $\Delta t$ are recovered (Fig.~\ref{fig:S10}). Over the random injections with $|\R|>0.1$, the median ratio of recovered to injected $|\R|$ is $0.995$ ($1.000$ with the DW fitted), and for $|\R|>0.1$ the delay is recovered within $0.5$\,ms in $98\%$ of cases.
 
\begin{figure}[!htb]
\centering\includegraphics[scale=0.45]{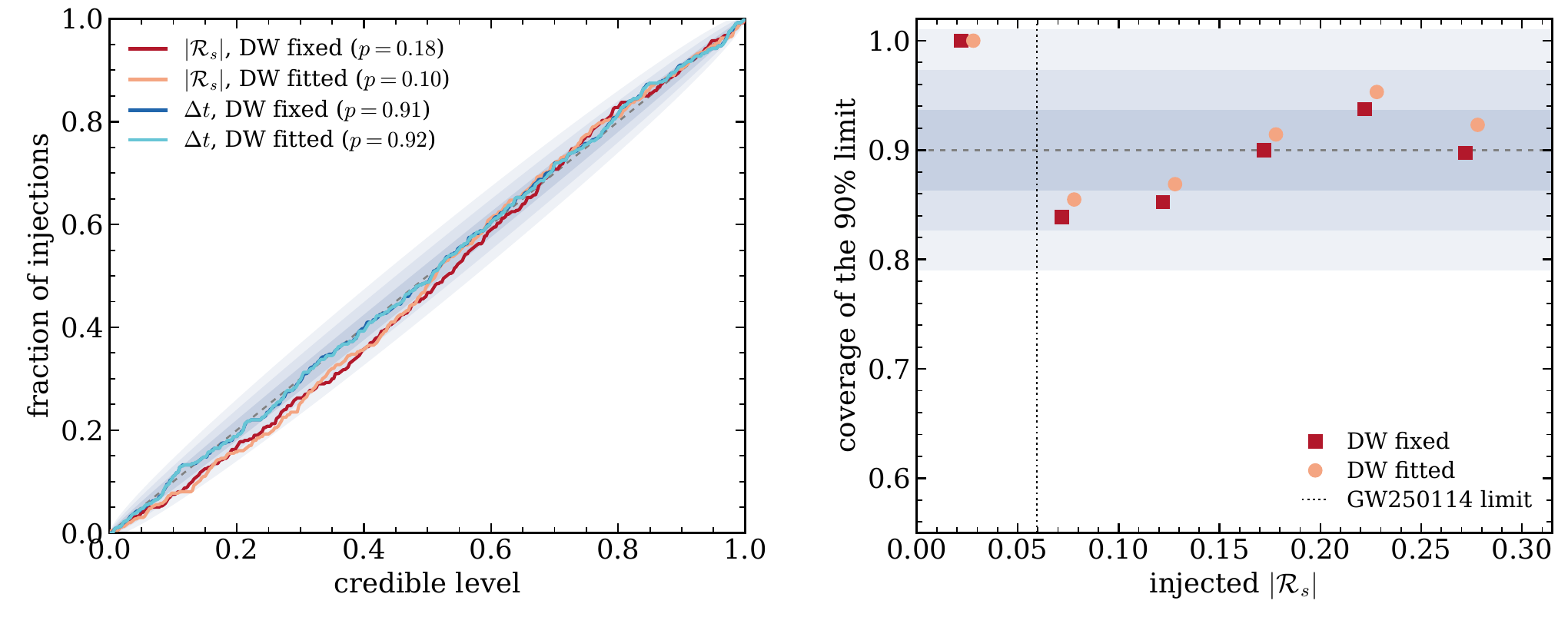}
\caption{Calibration with $400$ random injections, with the DW held fixed (dark) and fitted together with the echo (light). Left: fraction of injections whose true $|\R|$ (red) or $\Delta t$ (blue) lies below a given credible level, with the Kolmogorov--Smirnov $p$-values; shaded: $1\sigma$--$3\sigma$ ranges for a calibrated posterior. Right: fraction of injections, in bins of injected $|\R|$, with the 90\% limit above the injected $|\R|$; dotted: the GW250114 limit.}
\label{fig:S11}
\end{figure}

The credible levels of the injected $|\R|$ and $\Delta t$ are uniformly distributed (Kolmogorov--Smirnov $p=0.18$ and $0.91$, and $0.10$ and $0.92$ with the DW fitted; Fig.~\ref{fig:S11}, left), the recovered $|\R|$ shows no bias at any amplitude [Fig.~\ref{fig:constraint} (bottom)], and the 90\% limit lies above the injected $|\R|$ in $90.5\%$ of the injections ($92.0\%$ with the DW fitted; Fig.~\ref{fig:S11}, right). An echo above the quoted bound would therefore have been found, and the limit has its stated coverage.

\suppsection{Forecast for future detectors}
 
The echo SNR is $\rho=|\R|\,\rho_1$, with $\rho_1^2=\sum_I4\int|\tilde z_I(f)|^2/S_I(f)\,df$ for the template at $\R=1$, averaged over $\phi_R$. In Gaussian noise the 90\% limit in the absence of an echo scales as $1/\rho_1$, so the limit expected from another network is the O4 median off-source limit, $0.063$, divided by $\rho_1/\rho_1^{\rm O4}$. We keep the source of GW250114 fixed (distance, sky position and orientation) and evaluate $\rho_1$ with the design noise curves of A+, the Einstein Telescope (10-km triangle) and Cosmic Explorer (40 and 20\,km), as distributed with \texttt{gwbench} \cite{Borhanian:2020ypi_sup}; detector locations are not modeled, which changes the numbers by tens of percent. For the O4 detectors this procedure reproduces the $\rho_1$ of the search. Next-generation networks gain an order of magnitude in $|\R|$ and two in the reflected energy. These are statistical limits: at this sensitivity the template systematics of Sec.~II, not the noise, may set the accuracy of the bound.

\suppsection{Reflectivity and Echoes from ECOs in the Teukolsky framework}
\label{sec:SM_green}

This supplemental section collects results from the literature on which the ECO waveform model of the main text is based, viz., the Green's functions of the radial Teukolsky equation for a Kerr BH and ECO, the construction of the ECO waveform, and the subtleties involved in imposing reflective boundary conditions at the ECO surface within the Teukolsky framework. We include it to keep the paper self-contained and because the references on which it is based use different notation and conventions. Therefore, we recast the results here in the notation of the main text. The construction of the Green's function follows Refs.~\cite{Mark:2017dnq_sup,Motohashi:2026mbn_sup}, and the boundary condition is based on the membrane paradigm approach of Refs.~\cite{Chen:2020htz_sup,Xin:2021zir_sup,Ma:2022wpv_sup}. Along the way, we sketch how to derive the echo amplitude given by Eq.~\eqref{eq:Z1} of the main text. A detailed review of the Teukolsky framework can be found in \cite{Frolov98_sup,Sasaki:2003xr_sup}.

We consider a Kerr background of mass $M$ and spin parameter $a=\chi M$, with $\chi$ denoting the dimensionless spin parameter. We work with the spin weight $s=-2$, for which the Teukolsky master variable is $\rho^{-4}\psi_4$, with $\rho=-(r-ia\cos\theta)^{-1}$; the master variable for $s=+2$ is $\psi_0$. Both Newman--Penrose scalars, $\psi_0$ and $\psi_4$, are defined in the Kinnersley tetrad. We use the convention $e^{-i\omega t}$ in the Fourier decomposition of the modes and suppress the mode indices $(\ell,m,\omega)$, where $\ell$ and $m$ are the multipole and azimuthal indices. The horizon radius is $r_+=M+\sqrt{M^2-a^2}$, the horizon angular velocity is $\Omega_H=a/(2Mr_+)$, the surface gravity is $\kappa=\sqrt{M^2-a^2}/(2Mr_+)$. Lastly, $k_H=\omega-m\Omega_H$ is the frequency in the frame corotating with the horizon. Note that, in the limit $r \to r_+,$ $\Delta \equiv r^2-2Mr+a^2\simeq e^{2\kappa r_*}$.

\begin{figure}[htbp!]
    \centering
    \includegraphics[width=0.98\textwidth]
    {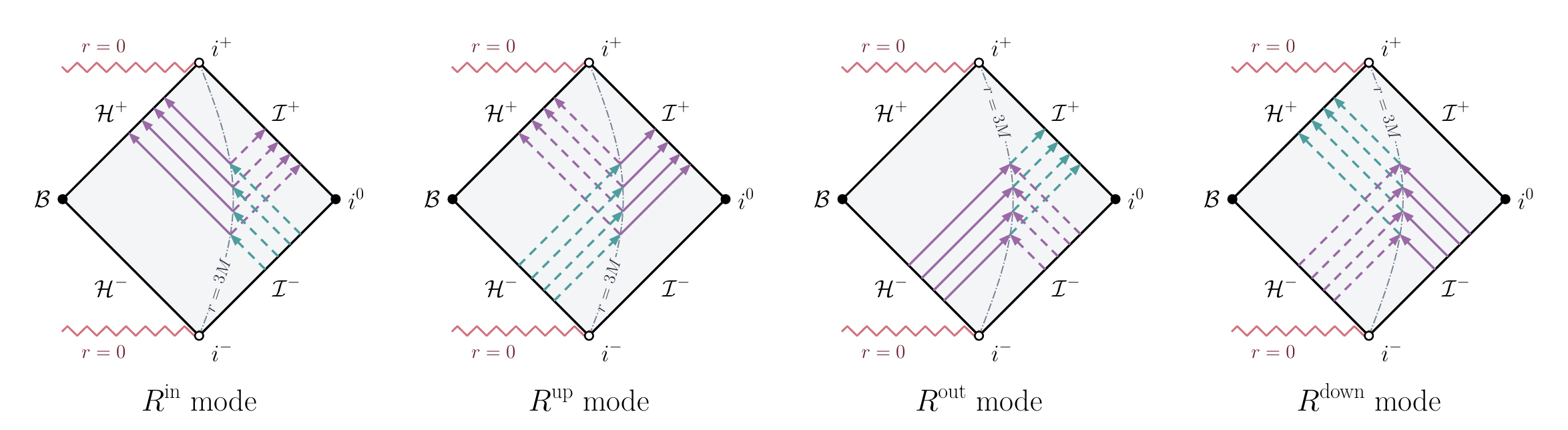}
    \caption{Penrose diagrams of the four homogeneous solutions of the radial equation, defined up to normalization by their boundary conditions: purely ingoing at $\mathcal{H}^+$ ($R^{\rm in}$), purely outgoing at $\mathcal{I}^+$ ($R^{\rm up}$), purely outgoing at $\mathcal{H}^-$ ($R^{\rm out}$), and purely ingoing at $\mathcal{I}^-$ ($R^{\rm down}$). The arrows show, in the eikonal limit, the scattering of the waves off the potential barrier near the light ring (gray dash-dotted worldline, $r=3M$). As a mnemonic, each name can be read off from the side of the barrier that carries a single wave (solid line): on the horizon side, this wave either enters $\mathcal{H}^+$ (in) or emerges from $\mathcal{H}^-$ (out); on the asymptotically flat side, it either reaches $\mathcal{I}^+$ (up) or arrives from $\mathcal{I}^-$ (down). The diagrams are drawn for Schwarzschild; for the $s=-2$ Teukolsky equation in Kerr they are schematically the same.}
    \label{fig:homogen_sols}
\end{figure}

\emph{\textbf{BH Green's function.---}}The radial function obeys the Teukolsky equation for spin weight $s=-2$ with a source term~\cite{Mino:2008at_sup}
\begin{equation}
    \Delta^{2}\frac{d}{dr}\!\left(\frac{1}{\Delta}\frac{dR}{dr}\right)+V(r)\,R=\mathcal{T}(r),
    \label{eq:SM_teuk}
\end{equation}
with $V=[K^2+4i(r-M)K]/\Delta-8i\omega r-\lambda$, where $K=(r^2+a^2)\omega-ma$ and $\lambda$ is the separation constant. The source $\mathcal{T}$ is built from the stress--energy tensor of the plunging particle~\cite{Mino:2008at_sup,Xin:2021zir_sup}. The two homogeneous solutions of Eq.~\eqref{eq:SM_teuk}, $R^{\rm in}$ and $R^{\rm up}$, have the following well-known asymptotic behavior~\cite{Mino:2008at_sup},
\begin{equation}
\begin{aligned}
R^{\rm in}&\to\begin{cases}B^{\rm trans}\Delta^2e^{-ik_Hr_*}, & r\to r_+,\\ B^{\rm ref}r^3e^{i\omega r_*}+B^{\rm in}e^{-i\omega r_*}/r, & r\to\infty,\end{cases}\\
R^{\rm up}&\to\begin{cases}C^{\rm up}e^{ik_Hr_*}+C^{\rm ref}\Delta^2e^{-ik_Hr_*}, & r\to r_+,\\ C^{\rm trans}r^3e^{i\omega r_*}, & r\to\infty.\end{cases}
\end{aligned}
\label{eq:asym}
\end{equation}
The radial equation also admits the two homogeneous solutions $R^{\rm out}$ and $R^{\rm down}$, which are purely outgoing at the horizon, $R^{\rm out}\to e^{ik_Hr_*}$, and purely ingoing at infinity, $R^{\rm down}\to e^{-i\omega r_*}/r$~\cite{Motohashi:2026mbn_sup}.\footnote{Since Eq.~\eqref{eq:SM_teuk} is of second order, only two of the four solutions are linearly independent. The pairs $\{R^{\rm in},R^{\rm out}\}$ and $\{R^{\rm up},R^{\rm down}\}$, adapted to the horizon and to infinity, respectively, are independent for all $\omega\neq0$, and each solution can be expanded in either pair; e.g., from Eq.~\eqref{eq:asym}, $R^{\rm in}=(B^{\rm ref}/C^{\rm trans})\,R^{\rm up}+B^{\rm in}R^{\rm down}$ and $R^{\rm up}=C^{\rm up}R^{\rm out}+(C^{\rm ref}/B^{\rm trans})\,R^{\rm in}$. The pair $\{R^{\rm in},R^{\rm up}\}$ is independent whenever $B^{\rm in}\neq0$, i.e., everywhere except at the quasinormal frequencies, where $R^{\rm in}\propto R^{\rm up}$ and the Wronskian below vanishes. Each of $R^{\rm in}$ and $R^{\rm up}$ satisfies one of the physical boundary conditions: nothing comes out of the past horizon, and nothing comes in from past null infinity either. Therefore this set gives the right ingredients for constructing the retarded Green's function in Eq.~\eqref{eq:SM_gBH}. On the other hand $\{R^{\rm out},R^{\rm down}\}$ has the time-reversed conditions (Fig.~\ref{fig:homogen_sols}).} The causal structure of all four solutions is illustrated in Fig.~\ref{fig:homogen_sols}; in what follows, we work with $R^{\rm in}$ and $R^{\rm up}$.

Now, the Wronskian for Eq.~\eqref{eq:SM_teuk} can be constructed using the $R^{\rm in}$ and $R^{\rm up}$ solutions, i.e., $W=R^{\rm in}R^{\rm up\,\prime}-R^{\rm up}R^{\rm in\,\prime}$. The quantity $W/\Delta$ turns out to be a constant instead of $W$ itself. We can evaluate it as $r\to\infty$ to get
\begin{equation}
    \frac{W}{\Delta}=2i\omega\,B^{\rm in}C^{\rm trans}.
\end{equation}
The Kerr BH Green's function, that is ingoing at the horizon and outgoing at infinity, is
\begin{equation}
    g_{\rm BH}(r,r')=\frac{R^{\rm in}(r_<)\,R^{\rm up}(r_>)}{2i\omega\,B^{\rm in}C^{\rm trans}\,\Delta^2(r')},
    \label{eq:SM_gBH}
\end{equation}
with $r_<=\min(r,r')$ and $r_>=\max(r,r')$. One can write down the normalization from the jump condition $[\,\partial_r g_{\rm BH}\,]_{r'^-}^{r'^+}=1/\Delta(r')$, obtained by integrating Eq.~\eqref{eq:SM_teuk} across $r=r'$. The solution is then given by $R(r)=\int dr'\,g_{\rm BH}(r,r')\,\mathcal{T}(r')$, and its amplitudes at infinity and at the horizon are
\begin{equation}
    Z^\infty=\frac{1}{2i\omega B^{\rm in}}\int dr'\,\frac{R^{\rm in}(r')\,\mathcal{T}(r')}{\Delta^2(r')},
    \qquad
    Z^H=\frac{B^{\rm trans}}{2i\omega B^{\rm in}C^{\rm trans}}\int dr'\,\frac{R^{\rm up}(r')\,\mathcal{T}(r')}{\Delta^2(r')},
    \label{eq:SM_ZH}
\end{equation}
following the usual definitions, viz., $R\to Z^\infty r^3e^{i\omega r_*}$ as $r\to\infty$ and $R\to Z^H\Delta^2e^{-ik_Hr_*}$ as $r\to r_+$. As stated in the main text, each amplitude is a source integral weighted by the homogeneous solution that satisfies the physical boundary condition at the opposite boundary.

\emph{\textbf{Ingoing and outgoing waves at the surface.---}}We place the reflecting surface at $r_*=r_*^s$ (radius $r_s$), close enough to the horizon such that $\Delta(r_s)\ll M^2$. Between the surface and the source located outside, the $s=-2$ radial function is a superposition of the two near-horizon solutions of Eq.~\eqref{eq:asym},
\begin{equation}
    R\simeq c_{\rm in}\,\Delta^2e^{-ik_Hr_*}+c_{\rm out}\,e^{ik_Hr_*},
    \label{eq:SM_surface}
\end{equation}
where $c_{\rm in}$ is the amplitude of the wave traveling into the surface and $c_{\rm out}$ is the amplitude of the wave leaving it. For a Kerr BH, $c_{\rm out}=0$ and $c_{\rm in}=Z^H$; but in the presence of a reflecting surface, $c_{\rm in}$ also contains the part of the reflected wave that the potential barrier returns to the surface.

The two terms appearing in Eq.~\eqref{eq:SM_surface} are not of the same order since the ingoing wave present in the $s=-2$ field is suppressed by a factor of $\Delta^2$. In the Kinnersley tetrad, $\psi_4$ is the radiative Weyl scalar for outgoing waves. But near the surface, the ingoing wave is carried at leading order by
\begin{equation}
    \psi_0\simeq d_{\rm in}\,\Delta^{-2}e^{-ik_Hr_*},
    \label{eq:SM_psi0}
\end{equation}
where $d_{\rm in}$ is the ingoing amplitude of the $s=+2$ radial function, which near the surface behaves as $d_{\rm in}\Delta^{-2}e^{-ik_Hr_*}+d_{\rm out}\,e^{ik_Hr_*}$. This radial function is the $s=+2$ counterpart of the $s=-2$ radial function in Eq.~\eqref{eq:SM_surface}.

We cannot define a reflective boundary condition at the ECO surface using the $s=-2$ field alone. To see this, note that a finite reflectivity $\mathcal{R}'$ imposed on the $\psi_4$ field at the surface gives $c_{\rm out}e^{ik_Hr_*^s}=\mathcal{R}'c_{\rm in}\Delta^2(r_s)e^{-ik_Hr_*^s}$, which in turn implies,
\begin{equation}
    \frac{c_{\rm out}}{c_{\rm in}}=\mathcal{R}'\,\Delta^2(r_s)\,e^{-2ik_Hr_*^s},
    \qquad
    \left|\frac{c_{\rm out}}{c_{\rm in}}\right|\underset{r_s \to r_+}{\propto} |\mathcal R'|e^{4\kappa r_*^s},
\end{equation}
which vanishes as $r_s\to r_+$. Therefore such a surface would apparently reflect nothing back, regardless of the value of $|\mathcal{R}'|$. Moreover, as shown below, $|c_{\rm out}/c_{\rm in}|^2$ is not the reflected fraction of the energy flux, so the ratio does not define a physical reflectivity. Hence one cannot naively impose reflective boundary conditions on the ECO surface in this fashion. A physical reflectivity must compare the incident $\psi_0$ with the reflected $\psi_4$. The tidal field measured by observers hovering near the surface can be used to obtain this ratio. The Teukolsky--Starobinsky (TS) identity can then be used to convert the result back into a condition on $c_{\rm in}$ and $c_{\rm out}$.

\emph{\textbf{Membrane paradigm boundary condition.---}}The response of the ECO is defined in the frame of zero-angular-momentum fiducial observers (FIDOs) hovering just above the would-be horizon, with four-velocity $U^b$ and spatial projector $h_i^a=\delta_i^a+U^aU_i$~\cite{Chen:2020htz_sup}. These observers measure the electric tidal tensor $\mathcal{E}_{ij}=h_i^ah_j^cC_{abcd}U^bU^d$, where $C_{abcd}$ is the Weyl tensor. Near the horizon, its transverse component is~\cite{Chen:2020htz_sup}
\begin{equation}
    \mathcal{E}_{mm}\simeq-\frac{\Delta}{4\Sigma}\,\psi_0-\frac{\Sigma}{\Delta}\,\psi_4^*,
    \qquad \Sigma=r^2+a^2\cos^2\theta.
    \label{eq:Emm}
\end{equation}
The ingoing part of $\psi_0$ scales as $\Delta^{-2}$ and the outgoing part of $\psi_4$ as $\Delta^{0}$, so both terms diverge as $\Delta^{-1}$, indicating the blueshift of the waves measured by the FIDOs. The outgoing part of $\psi_0$ and the ingoing part of $\psi_4$ enter only at $\mathcal{O}(\Delta)$ and are negligible. Using $\psi_4^*=\rho^{*4}(\rho^{-4}\psi_4)^*$, the outgoing wave $\rho^{-4}\psi_4\simeq c_{\rm out}e^{ik_Hr_*}$, and the ingoing wave given by~Eq.~\eqref{eq:SM_psi0}, the incident and response (outgoing) fields in $\mathcal{E}_{mm}$  can be written as (see Eqs.~(21)--(24) of Ref.~\cite{Chen:2020htz_sup})
\begin{equation}
    \mathcal{E}^{\rm src}_{mm}\simeq-\frac{d_{\rm in}\,e^{-ik_Hr_*}}{4\Sigma\Delta},
    \qquad
    \mathcal{E}^{\rm resp}_{mm}\simeq-\frac{\Sigma\rho^{*4}}{\Delta}\bigl(c_{\rm out}e^{ik_Hr_*}\bigr)^*,
    \label{eq:SM_Esplit}
\end{equation}
where we suppress the angular and time dependence. Within linear response theory, the ECO responds to the total tidal field at its surface, that is, $\mathcal{E}^{\rm resp}_{mm}=\eta\mathcal{E}_{mm}=\eta\,(\mathcal{E}^{\rm src}_{mm}+\mathcal{E}^{\rm resp}_{mm})$, where $\eta$ is a tidal Love-number-like coefficient. This enables us to define the surface reflectivity as
\begin{equation}
    \mathcal{E}^{\rm resp}_{mm}=-\R\,\mathcal{E}^{\rm src}_{mm}\,, 
    \label{eq:SM_Rdef}
\end{equation}
where $\R=-{\eta}/{(1-\eta)}$~\cite{Chen:2020htz_sup}. The common factor $\Delta^{-1}$ in Eq.~\eqref{eq:SM_Esplit} cancels, so $\R$ is finite, independent of $\Delta$, and relates the finite amplitudes $c_{\rm out}$ and $d_{\rm in}$. Its form is set by the physics of the surface, e.g., the Boltzmann or Lorentzian reflectivities~\cite{Xin:2021zir_sup}. It is evident that $\R=0$ corresponds to a Kerr BH, and $|\R|=1$ is a perfectly reflecting ECO.

\emph{\textbf{Physical Teukolsky reflectivity.---}}Since $\psi_4$ enters Eq.~\eqref{eq:Emm} in its complex conjugate form, the condition~\eqref{eq:SM_Rdef} couples the $(\ell,m,\omega)$ mode of $\psi_4$ to the $(\ell,-m,-\omega^*)$ mode of $\psi_0$, and the $\theta$ dependence of the surface response and of the tetrad couples different $\ell$-modes as well~\cite{Chen:2020htz_sup}. We shall neglect this so-called $\ell$-mixing. For a source symmetric under reflection through the orbital plane~\cite{Boyle:2014ioa_sup}, such as the equatorial plunge, the $(\ell,-m,-\omega^*)$ amplitude equals $(-1)^\ell$ times the complex conjugate of the $(\ell,m,\omega)$ amplitude. Imposing Eq.~\eqref{eq:SM_Rdef} at $r_*=r_*^s$ then gives the following relation~\cite{Chen:2020htz_sup}
\begin{equation}
    c_{\rm out}\,e^{ik_Hr_*^s}=\frac{(-1)^{\ell+m+1}}{4}\,\R\,d_{\rm in}\,e^{-ik_Hr_*^s}.
    \label{eq:SM_local}
\end{equation}
The overall sign collects the minus sign of the response relation, a factor $(-1)^m$ from the spin-weighted harmonics, and the factor $(-1)^\ell$ of the reflection symmetry. The factor $1/4$ comes from the modulus of $1/(4\Sigma^2\rho^{*4})$ in Eq.~\eqref{eq:SM_Esplit}, whose $\theta$-dependent phase has been discarded since we choose to neglect $\ell$-mixing.\footnote{Since $\rho^*=-(r+ia\cos\theta)^{-1}$ and $\Sigma=(r-ia\cos\theta)(r+ia\cos\theta)$, one has $\Sigma^2\rho^{*4}=[(r-ia\cos\theta)/(r+ia\cos\theta)]^2$, such that $|1/(4\Sigma^2\rho^{*4})|=1/4$. At $r_s\simeq r_+$, its phase is $e^{-4i \tilde \beta(\theta)}$, with $ \tilde \beta=-\arctan(a\cos\theta/r_+)$~\cite{Chen:2020htz_sup}. The phase is unity only for $a=0$ or on the equator. Since the phase is $\theta$ dependent, it couples different $\ell$ when projected onto spin-weighted spheroidal harmonics. So, taking an $\R$ independent of $\theta$ and discarding this phase together amounts to neglecting the $\ell$-mixing.} The modulus of Eq.~\eqref{eq:SM_local} is therefore $|c_{\rm out}|=|\R|\,|d_{\rm in}|/4$, and one can then show that $|\R|^2$ indeed denotes the ratio of the reflected to the incident energy flux at the surface (see Appendix~B of Ref.~\cite{Chen:2020htz_sup} for details).

The TS identity can now be used to relate the ingoing amplitudes of the two spin fields, $d_{\rm in}=(D_{\ell m}/C_{\ell m})\,c_{\rm in}$~\cite{Chen:2020htz_sup}, and eliminate $d_{\rm in}$ from Eq.~\eqref{eq:SM_local}. We therefore get
\begin{equation}
    c_{\rm out}\,e^{ik_Hr_*^s}=\hat{\sigma}\,\R\,c_{\rm in}\,e^{-ik_Hr_*^s},
    \qquad
    \hat{\sigma}\equiv\frac{(-1)^{\ell+m+1}}{4}\,\frac{D_{\ell m}}{C_{\ell m}},
    \label{eq:SM_local2}
\end{equation}
which for the $\ell=2$ mode considered here is $\hat{\sigma}=(-1)^{m+1}D_{\ell m}/(4C_{\ell m})$. Note that
\begin{equation}
    D_{\ell m}=64(2Mr_+)^4(ik_H)\bigl(k_H^2+4\epsilon_\kappa^2\bigr)\bigl(-ik_H+4\epsilon_\kappa\bigr),
    \qquad
    \epsilon_\kappa=\frac{\kappa}{2}=\frac{\sqrt{M^2-a^2}}{4Mr_+},
    \label{eq:SM_D}
\end{equation}
is independent of $\ell$ and depends on $(m,\omega)$ only through $k_H$~\cite{Xin:2021zir_sup,Ma:2022wpv_sup}, and the Starobinsky constant $C_{\ell m}$ satisfies~\cite{Ma:2022wpv_sup}
\begin{equation}
\begin{aligned}
    |C_{\ell m}|^2 ={} & \bigl( Q^2 + 4a\omega m - 4a^2\omega^2 \bigr)\bigl[ (Q-2)^2 + 36a\omega m - 36a^2\omega^2 \bigr]\\
    & + (2Q-1)(96a^2\omega^2 - 48a\omega m) + 144\omega^2(M^2-a^2),
\end{aligned}
\label{eq:SM_C}
\end{equation}
with $\Im\,C_{\ell m}=12M\omega$ and $\Re\,C_{\ell m}=+\sqrt{|C_{\ell m}|^2-144M^2\omega^2}$ where $Q=\lambda+2$, so that $Q\to\ell(\ell+1)$ as $a\omega\to0$.

Therefore, the membrane paradigm approach, along with the use of the TS identity, removes the difficulty we identified earlier while imposing a reflectivity on the $s=-2$ field alone. 

Now, $|c_{\rm out}/c_{\rm in}|^2=|\hat{\sigma}|^2|\R|^2$. Clearly, it no longer carries the irksome $\Delta^2$ factor and it differs from the reflected energy fraction $|\R|^2$ by the frequency-dependent factor $|\hat{\sigma}|^2$.
A reflectivity imposed directly on $c_{\rm out}/c_{\rm in}$ would omit this factor. Moreover, note that from Eq.~\eqref{eq:SM_local2}, $\hat{\sigma}\R$ acquires a phase due to a round trip around the cavity,
\begin{equation}
    \frac{c_{\rm out}}{c_{\rm in}}=\hat{\sigma}\,\R\,e^{-2ik_Hr_*^s}=\hat{\sigma}\,\R\,e^{ik_H\Delta t}\equiv\hat{\mathcal{G}},
    \qquad
    \Delta t=-2r_*^s,
    \label{eq:SM_bc}
\end{equation}
which is the factor $\hat{\mathcal{G}}$ of Eq.~\eqref{eq:Z1} (cf.\ Eq.~\eqref{eq:dt}). Because $\Im\,C_{\ell m}=12M\omega$, $C_{\ell m}\neq0$ for real $\omega\neq0$, while on the real axis $D_{\ell m}$ vanishes only at $k_H=0$. Hence $\hat{\mathcal{G}}$ is finite and, for $k_H\neq0$, nonzero. It vanishes linearly at the superradiant threshold $k_H=0$, consistent with the vanishing of the echo at $f=m\Omega_H/2\pi$ (Fig.~\ref{fig:template}). We also take $\R=|\R|e^{i\phi_{{R}}}$ to be constant over the analysis band.\footnote{Technically, the condition~\eqref{eq:SM_Rdef} results in the complex conjugate of $\R$ evaluated at the frequency $-\omega^*+m\Omega_H$ appearing in Eq.~\eqref{eq:SM_local}; but we still denote it by $\R$, which leaves $|\R|$ unchanged. The simplification requires the surface to preserve the reflection symmetry of the source, i.e., a real time-domain response, $\R(-k_H)=\R(k_H)^*$. For constant $\R$, this holds only for $\phi_{\mathcal{R}}\in\{0,\pi\}$ (see Ref.~\cite{Chen:2020htz_sup} for details).}

\emph{\textbf{ECO Green's function.---}}The ECO Green's function $g_{\rm ECO}$ satisfies the same equation as $g_{\rm BH}$, so their difference is a homogeneous solution; since it must be outgoing at infinity, it is proportional to $R^{\rm up}$,
\begin{equation}
    g_{\rm ECO}(r,r')=g_{\rm BH}(r,r')+\mathcal{A}(r')\,R^{\rm up}(r).
\end{equation}
For a source point $r'$ outside the surface and $r_s<r<r'$, $g_{\rm ECO}(r,r')$ obeys the boundary condition given by Eq.~\eqref{eq:SM_surface} with $r'$-dependent coefficients. We can use Eqs.~\eqref{eq:asym} and~\eqref{eq:SM_gBH} to then get
\begin{equation}
    c_{\rm in}(r')=\frac{B^{\rm trans}R^{\rm up}(r')}{2i\omega B^{\rm in}C^{\rm trans}\Delta^2(r')}+\mathcal{A}(r')\,C^{\rm ref},
    \qquad
    c_{\rm out}(r')=\mathcal{A}(r')\,C^{\rm up}.
\end{equation}
Using Eq.~\eqref{eq:SM_bc} we can determine $\mathcal{A}(r')$. In terms of the barrier reflection and transmission amplitudes for a wave leaving the surface, that is, $r^{\rm BH}=C^{\rm ref}/C^{\rm up}$ and $t^{\rm BH}=C^{\rm trans}/C^{\rm up}$, we can write
\begin{equation}
    g_{\rm ECO}(r,r')=g_{\rm BH}(r,r')
    +\frac{B^{\rm trans}\,\mathcal{K}}{2i\omega B^{\rm in}(C^{\rm trans})^2}\,
    \frac{R^{\rm up}(r)\,R^{\rm up}(r')}{\Delta^2(r')},
    \label{eq:SM_gECO}
\end{equation}
where
\begin{equation}
    \mathcal{K}=\frac{t^{\rm BH}\,\hat{\mathcal{G}}}{1-r^{\rm BH}\,\hat{\mathcal{G}}}
    \label{eq:SM_K}
\end{equation}
is called the transfer function. The Kerr BH Green's function is recovered for $\R\to0$.

\emph{\textbf{ECO waveform.---}}Inserting Eq.~\eqref{eq:SM_gECO} into $R(r)=\int dr'\,g_{\rm ECO}(r,r')\,\mathcal{T}(r')$ and using $R^{\rm up}\to C^{\rm trans}r^3e^{i\omega r_*}$ as $r\to\infty$, one factor of $C^{\rm trans}$ cancels out and the second term reduces to $\mathcal{K}Z^H$ (Eq.~\eqref{eq:SM_ZH}), so that the amplitude at infinity is given by
\begin{equation}
    Z^\infty_{\rm ECO}=Z^\infty+\mathcal{K}\,Z^H.
    \label{eq:SM_ZECO}
\end{equation}
The ECO waveform is thus the BH waveform plus the horizon amplitude $Z^H$ filtered by the transfer function $\mathcal{K}$~\cite{Mark:2017dnq_sup}. In other words, the radiation that a BH would have absorbed is partly reflected by the surface and transmitted through the barrier to infinity. Expanding $\mathcal{K}$ in powers of $r^{\rm BH}\hat{\mathcal{G}}$ gives a sum over round trips in the cavity. Its first term, $\hat{\mathcal{G}}\,t^{\rm BH}Z^H$, is the first echo of Eq.~\eqref{eq:Z1}, which is the only term used in our search. We write $r^{\rm BH}\hat{\mathcal{G}}=\R\,\mathcal{R}^{\rm BH}e^{ik_H\Delta t}$, where $\mathcal{R}^{\rm BH}=\hat{\sigma}\,r^{\rm BH}$ is the flux-normalized reflectivity of the potential barrier~\cite{Xin:2021zir_sup,Rosato:2026apq_sup}. Indeed, $|\mathcal{R}^{\rm BH}|^2$ is the fraction of the energy flux leaving the surface that the barrier returns to it, and $\Gamma=1-|\mathcal{R}^{\rm BH}|^2$ is the barrier transmissivity. The expansion converges for $|\R\,\mathcal{R}^{\rm BH}|<1$. In the superradiant regime $0<\omega<m\Omega_H$, $|\mathcal{R}^{\rm BH}|>1$ and $\Gamma<0$, so stability of the ECO requires $|\R\,\mathcal{R}^{\rm BH}|<1$ in this frequency range~\cite{Xin:2021zir_sup}.

\emph{\textbf{Limitations.---}}The TS identity strictly holds only for homogeneous solutions of the Teukolsky equation. With the source being a plunging particle, the conversion between the ingoing $\psi_4$ and $\psi_0$ may require further study~\cite{Xin:2021zir_sup,Srivastava:2021uku_sup}. 
We have also neglected the $\ell$-mixing induced by the spin of the ECO and reflections from its surface~\cite{Chen:2020htz_sup}. Finally, the membrane paradigm boundary condition assumes a local, linear tidal response of the surface; more general scenarios, such as feedback terms in the Teukolsky equation or a nonlinear response of the near-horizon region rather than the linear partial reflection considered here, may require a modified prescription~\cite{Xin:2021zir_sup,Deppe:2025pvd_sup,Srivastava:2021uku_sup,Jiang:2025mna_sup,Lo:2025lpo_sup,Yin:2025kls_sup}.

\putbib[sup_bibliography]
\end{bibunit}


\begin{thebibliography}{73}
\bibitem [{Not({\natexlab{a}})}]{NotationConventions}%
  \BibitemOpen
  \bibinfo {note} {Notation and conventions: We use $G=c=1$ and the mostly-plus signature. For a Kerr BH, $a=J/M$, $\chi=a/M$, and $r_\pm=M\pm\sqrt{M^2-a^2}$, where the symbols have their usual meanings unless stated otherwise. The ECO surface lies at $r_s=r_++\epsilon$. For the remnant, we use $M=\Mf$ and $\chi=\chi_f$.}\BibitemShut {Stop}%
\end{thebibliography}
\end{document}